\documentclass[fleqn,usenatbib]{mnras}

\usepackage{graphicx}
\usepackage{amsmath}
\usepackage{amssymb}
\usepackage{booktabs}
\usepackage{bm}
\usepackage{parskip}

\usepackage{microtype}
\usepackage{tikz}
\usetikzlibrary{automata, positioning, arrows.meta, decorations.pathreplacing}

\definecolor{sym0}{HTML}{1565C0} 
\definecolor{sym1}{HTML}{C62828} 
\definecolor{sym2}{HTML}{2E7D32} 
\definecolor{sym3}{HTML}{E65100} 
\newcommand{\tl}[2]{\ensuremath{\textcolor{sym#1}{#1}{:}#2}}

\graphicspath{{figures/}}

\newcommand{\Cmu}{C_\mu}
\newcommand{\hmu}{h_\mu}
\newcommand{\emachine}{\ensuremath{\varepsilon}-machine}
\newcommand{\emachines}{\ensuremath{\varepsilon}-machines}
\newcommand{\bits}{\,\mathrm{bits}}
\newcommand{\bitssym}{\,\mathrm{bits\,sym^{-1}}}
\newcommand{\frba}{FRB~20121102A}
\newcommand{\frbb}{FRB~20201124A}
\newcommand{\frbc}{FRB~20220912A}

\title[Temporal Memory in Repeating FRBs]{Temporal Memory in Repeating Fast Radio Bursts: Epsilon-Machine Reconstruction of Causal Structure in Burst Timing}

\author[T. Kimpson and J. O'Leary]{
Tom Kimpson,$^{1}$\thanks{E-mail: tom.kimpson@unimelb.edu.au}
Joseph O'Leary$^{2,3}$
\\
$^{1}$School of Mathematics and Statistics, University of Melbourne, Parkville, VIC 3010, Australia\\
$^{2}$Innovative Planet Research Institute, Swinburne University of Technology, Hawthorn, VIC 3122, Australia\\
$^{3}$School of Physics, University of Melbourne, Parkville, VIC 3010, Australia
}

\date{Accepted XXX. Received YYY; in original form ZZZ}
\pubyear{2026}

\begin{document}
\label{firstpage}
\pagerange{\pageref{firstpage}--\pageref{lastpage}}
\maketitle

\begin{abstract}
The emission mechanism of fast radio bursts (FRBs) remains unknown. Whether the bursts from a repeating FRB arrive at random or in a structured sequence is a key constraint on that mechanism. We apply \emachine{} reconstruction, a tool from computational mechanics that infers the minimal model capturing all predictive information in a stochastic process. Applied to the waiting-time sequences of three repeating FRBs (\frba{} and \frbb{} from FAST; \frbc{} from CHIME), the method yields the statistical complexity $\Cmu$, the minimum number of bits required for optimal prediction. Both FAST sources carry roughly one bit of temporal memory (significant against permutation surrogates, $p \leq 0.01$; per-source false-discovery-rate-adjusted $p \leq 0.028$), while \frbc{} is consistent with memoryless emission. \frbb{}'s memory spans hours-to-days across four sessions, \frba{}'s spans hours-to-weeks across thirty-nine, and neither source shows defensible within-session predictive memory. For \frba{} the ordering of those sessions is itself predictive (session-shuffle $p = 0.02$), whereas \frbb{}'s signal reflects the contrast between heterogeneous sessions rather than their order. A simulated windowing test shows that CHIME's short transit observations would suppress comparable structure in the FAST data, leaving \frbc{}'s null result ambiguous. This first application of \emachine{} reconstruction to astrophysical transients yields a model-independent constraint: the bursting of at least two of these repeaters is not memoryless, but is governed by a hidden state that occupies distinct activity-rate regimes varying across observing sessions, behaviour that any viable physical model must reproduce. \vspace{5mm}
\end{abstract}


\section{Introduction}
\label{sec:intro}

Fast radio bursts (FRBs) are bright, millisecond-duration radio transients of predominantly extragalactic origin \citep{Lorimer2007, Petroff2022}. Since the discovery of the first repeating source FRB~20121102A \citep{Spitler2016}, the population of confirmed repeaters has grown to several dozen, yet the physical mechanism generating the repeating bursts remains an open question \citep{Platts2019, Zhang2023review}. The detection of an FRB-like burst from the Galactic soft gamma repeater SGR~1935+2154 \citep{Bochenek2020, CHIME2020nature} established magnetars as at least one class of FRB progenitor, but whether all repeaters share a common emission mechanism is unknown.

One of the central open questions for repeater models is whether the temporal sequence of bursts from a given source is structured or random \citep{Oppermann2018, Cruces2021, Zhang2024timeenergy, SangLin2024}. If past waiting times (the intervals between successive bursts) constrain future ones, this implies a hidden state (either intrinsic or extrinsic to the source) that evolves over time and modulates the probability of observing a burst. Conversely, if the waiting-time sequence is consistent with an independent and identically distributed (IID) process, the bursting at any given moment is memoryless and uncorrelated with the burst history. Distinguishing these scenarios constrains the space of viable physical models: a memoryless process is consistent with a stationary Poisson emitter, while structured timing demands an evolving hidden state.

Several approaches have been used to probe this question, each sensitive to a different aspect of the temporal structure. Waiting-time distribution analyses fit parametric models (exponential, Weibull, log-normal) to the observed waiting times \citep{Oppermann2018, Gourdji2019, Aggarwal2021, Cruces2021, Jahns2023}, testing whether the marginal distribution departs from the exponential form expected for a Poisson process; \citet{Li2021} additionally revealed a bimodal waiting-time distribution for \frba{} from observations with the Five-hundred-meter Aperture Spherical radio Telescope (FAST). Periodicity searches have revealed activity windows with periods of order days to weeks in some sources \citep{CHIME2020periodic, Rajwade2020}. More recently, \citet{Zhang2024timeenergy} applied approximate entropy \citep[the Pincus index;][]{Pincus1991} and maximum Lyapunov exponent to FRB arrival times and energies, finding that repeating FRBs behave more randomly and less chaotically than pulsars, earthquakes, or solar flares; \citet{SangLin2024} reached complementary conclusions for the FAST repeaters using the Pincus index, the Hurst exponent, and non-Gaussian fluctuation statistics. Related analyses likewise find that the waiting-time sequences of these FAST repeaters are not consistent with an independent random process, reporting correlations between successive intervals and clustering that persists across a wide range of timescales \citep{Du2024scaling, Wang2024memory, WangWuDai2023}.

Each of these methods captures a specific projection of the temporal structure, but none recovers the \emph{minimal causal model} generating the observed sequence. A Weibull fit characterises the shape of the waiting-time distribution but not whether consecutive waiting times are statistically dependent. A positive Lyapunov exponent indicates sensitivity to initial conditions but does not identify the hidden states driving it. What is missing is a framework that extracts the full predictive architecture of the process directly from data, without assuming a specific model form.

Computational mechanics provides such a framework. Developed by \citet{Crutchfield1989} and formalised by \citet{Shalizi2001}, it constructs the \emph{\emachine}: the unique minimal unifilar hidden Markov model (HMM) whose hidden states (the causal states) group together all histories that yield identical predictions about the future \citep{Upper1997, Crutchfield2012}. By `unifilar' we mean that the current hidden state and the next observation uniquely determine the next hidden state, so an observer who knows the current state never loses track of it; we give the formal definition in Section~\ref{sec:emachines}. For a stationary, ergodic process, the \emachine{} is provably the simplest model that captures all predictive information \citep{Shalizi2001}. Its Shannon entropy, known as the statistical complexity $\Cmu$, measures the minimum memory (in bits) that the process requires for optimal prediction. In astrophysics, the only prior application of \emachine{} reconstruction is the work of \citet{Bartlett2022}, who reconstructed \emachines{} from disc-integrated reflected light curves of Solar System planets and used $\Cmu$ as an agnostic biosignature, finding that Earth's statistical complexity exceeds Jupiter's by approximately 50\%. No application to astrophysical transients (FRBs, stellar flares, magnetar bursts, accretion variability) has been reported.

In this paper, we present the first application of \emachine{} reconstruction to astrophysical transient timing. We analyse the waiting-time sequences of three repeating FRB sources: \frba{} (FAST; \citealt{Li2021}), \frbb{} (FAST; \citealt{Zhang2022}), and \frbc{} (the Canadian Hydrogen Intensity Mapping Experiment, CHIME; \citealt{CHIME2026catalog2}). We reconstruct \emachines{} from symbolised waiting-time sequences using the Causal State Splitting Reconstruction (CSSR) algorithm \citep{Shalizi2004}, validate the pipeline on synthetic processes, and test the statistical significance of the results against permutation surrogates.

The remainder of this paper is organised as follows. Section~\ref{sec:compmech} provides a self-contained introduction to computational mechanics for an astrophysics audience. Section~\ref{sec:data} describes the data and preprocessing. Section~\ref{sec:methods} details the analysis methods. Section~\ref{sec:results} presents the main results. Section~\ref{sec:discussion} discusses the physical interpretation and limitations, and Section~\ref{sec:conclusions} summarises our conclusions.

\section{Computational Mechanics}
\label{sec:compmech}

This section provides a self-contained tutorial on computational mechanics for readers whose primary background is in observational astrophysics, following the pedagogical approach of \citet{Bartlett2022} and drawing on the foundational works of \citet{Crutchfield1989}, \citet{Shalizi2001}, and \citet{Crutchfield2012}.

\subsection{Predictive equivalence and causal states}
\label{sec:causal_states}

Consider a discrete-valued stochastic process $\{X_t\}_{t \in \mathbb{Z}}$, where each $X_t$ takes values in a finite alphabet $\mathcal{A}$. At any time $t$, the process has a past (the sequence of symbols already observed, $\overleftarrow{x} = \ldots, x_{t-2}, x_{t-1}$) and a future (the sequence yet to come, $\overrightarrow{x} = x_t, x_{t+1}, \ldots$). Different pasts may lead to different conditional distributions over futures. Two pasts $\overleftarrow{x}$ and $\overleftarrow{x}'$ are said to be \emph{predictively equivalent}, written $\overleftarrow{x} \sim_\varepsilon \overleftarrow{x}'$, if they yield the same conditional distribution:
\begin{equation}
    \label{eq:causal_states}
    \overleftarrow{x} \sim_\varepsilon \overleftarrow{x}' \iff P(\overrightarrow{X} \mid \overleftarrow{X} = \overleftarrow{x}) = P(\overrightarrow{X} \mid \overleftarrow{X} = \overleftarrow{x}').
\end{equation}
The equivalence classes under this relation are the \emph{causal states} $\mathcal{S} = \{s_1, s_2, \ldots\}$. These causal states form the coarsest partition of histories that preserves all predictive information about the future.

To ground this abstract setup in the problem at hand: for the FRB application developed below, each symbol $X_t$ in the sequence is one discretised waiting time between two consecutive bursts (the discretisation is described in Section~\ref{sec:symbolisation}), and the alphabet $\mathcal{A}$ is the small set of waiting-time bins (for instance, `short' versus `long'). The past $\overleftarrow{x}$ is the run of preceding waiting times and the future $\overrightarrow{x}$ the run still to come. A causal state is then the minimal summary of the past needed to predict the future: two histories of waiting times belong to the same causal state whenever they imply the same distribution over what follows.

\subsection{Epsilon-machines}
\label{sec:emachines}

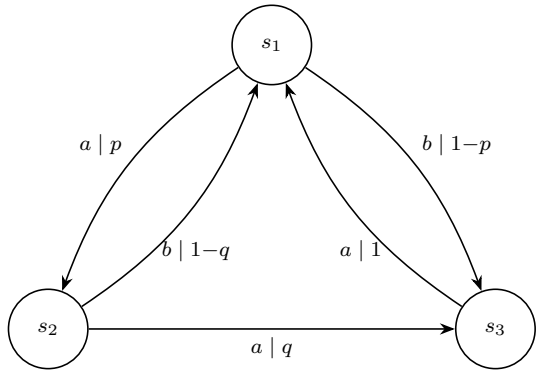
\begin{figure}
	\centering
	\begin{tikzpicture}[->, >=Stealth, semithick, node distance=3.0cm and 2.2cm,
		every state/.style={minimum size=1.05cm}]
		\node[state] (s1)                     {$s_1$};
		\node[state] (s2) [below left=of s1]  {$s_2$};
		\node[state] (s3) [below right=of s1] {$s_3$};
		\path (s1) edge [bend right=18] node[above left]     {$a \mid p$}     (s2)
		(s2) edge [bend right=18] node[pos=0.30, right] {$b \mid 1{-}q$} (s1)
		(s1) edge [bend left=18]  node[above right]    {$b \mid 1{-}p$} (s3)
		(s3) edge [bend left=18]  node[pos=0.30, left]  {$a \mid 1$}     (s1)
		(s2) edge                 node[below]           {$a \mid q$}     (s3);
	\end{tikzpicture}
	\caption{Schematic structure of a three-state \emachine. Circles are the causal states $s_i$, and each directed edge is labelled by an emitted symbol and its emission probability $P(x \mid s)$; the outgoing probabilities from each state sum to one. Unifilarity means that the current state and the emitted symbol together determine the next state uniquely (from $s_1$, for instance, $a$ leads to $s_2$ and $b$ to $s_3$), so all stochasticity is carried by the symbol emissions rather than by the state transitions. An \emachine{} may have any number of causal states; the three shown here, along with the transition topology and the values $(p, q)$, are purely illustrative.}
	\label{fig:abstract_emachine}
\end{figure}

\begin{figure}
	\centering
	\begin{tikzpicture}[font=\small]
		\node[anchor=west, font=\footnotesize\itshape] at (0,3.25) {Observed symbol sequence};
		\foreach \s [count=\i from 0] in {0,1,1,0,1,1,1,1,0,0,1,1} {
			\node[draw, minimum size=0.42cm, inner sep=0pt, font=\ttfamily\small]
				(d\i) at (\i*0.46,2.6) {\s};
		}
		\draw[decorate, decoration={brace, amplitude=4pt, mirror}, thin]
			(d4.south west) -- (d7.south east)
			node[midway, below, font=\scriptsize, yshift=-2pt] {even-length block of 1s};
		\draw[->, >=Stealth, semithick] (2.53,1.55) -- (2.53,0.95)
			node[midway, right, font=\footnotesize] {reconstruct \emachine};
		\begin{scope}[->, >=Stealth, semithick, every state/.style={minimum size=0.95cm}]
			\node[state] (eA) at (1.63,0.1) {$s_1$};
			\node[state] (eB) at (3.43,0.1) {$s_2$};
			\path (eA) edge [loop left]    node        {$0 \mid \tfrac{1}{2}$} (eA)
			      (eA) edge [bend left=22] node[above]  {$1 \mid \tfrac{1}{2}$} (eB)
			      (eB) edge [bend left=22] node[below]  {$1 \mid 1$}            (eA);
		\end{scope}
	\end{tikzpicture}
	\caption{The reconstruction task that motivates this paper, illustrated for the Even Process (1s occur only in even-length blocks, bounded by 0s). The observable is a symbol sequence (top); what we infer from it is the underlying \emachine{} (bottom). State $s_1$ is stochastic, emitting ``0'' or ``1'' with equal probability, and state $s_2$ is deterministic, emitting the ``1'' that completes an even block; edges are labelled by emitted symbol and emission probability $P(x \mid s)$. The hidden causal state is never observed directly: it is reconstructed from the statistics of the sequence, here by the CSSR algorithm (Section~\ref{sec:cssr_method}). This Even Process \emachine{} is one of the analytically known test cases used to validate the pipeline in Appendix~\ref{sec:known_machines}.}
	\label{fig:reconstruction}
\end{figure}
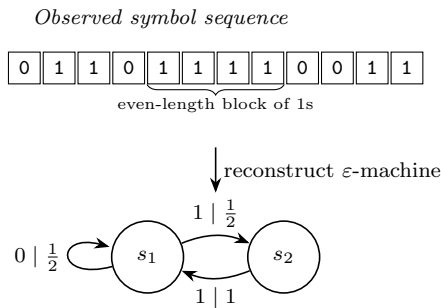

The \emachine{} is the stochastic automaton whose internal states are the causal states. At each time step, the machine occupies a causal state $s$ and emits a symbol $x \in \mathcal{A}$ with probability $P(X_t = x \mid S_t = s)$. The observed symbol $X_t$ is the quantity we measure (here, the next waiting time), whereas the causal state $S_t$ is hidden and inferred from the observed sequence; $s$ denotes one particular value of that hidden state. The defining property that distinguishes the \emachine{} from a generic hidden Markov model is \emph{unifilarity}: once the current state $s$ and the emitted symbol $x$ are known, the next state $s' = \delta(s, x)$ is fixed. There is no residual stochasticity in the state transition itself; all the randomness in the process lives in the symbol emission. Figure~\ref{fig:abstract_emachine} shows the general structure of an \emachine{} schematically, and Figure~\ref{fig:reconstruction} illustrates the reconstruction task that the rest of this paper performs: recovering such a machine from an observed symbol sequence.

\citet{Shalizi2001} proved three results about \emachines. First, every stationary ergodic process has a unique \emachine. Second, the \emachine{} is minimal: no other unifilar HMM of the process has fewer states. Third, the \emachine{} is sufficient: it captures all of the mutual information between the past and the future. Together, these properties single out the \emachine{} among models of the process: it is the simplest model that loses no predictive information.

\subsection{Key information-theoretic measures}
\label{sec:measures}

The \emachine{} defines three information-theoretic quantities standard in computational mechanics:

\paragraph{Statistical complexity $\Cmu$.} The Shannon entropy \citep{Shannon1948, CoverThomas2006} of the stationary distribution $\pi$ over causal states. Writing $S$ for the stationary causal-state random variable,
\begin{equation}
    \Cmu = H[S] = -\sum_{s \in \mathcal{S}} \pi(s) \log_2 \pi(s).
    \label{eq:cmu}
\end{equation}
Equation~(\ref{eq:cmu}) defines the quantity we report throughout this paper. $\Cmu$ measures the minimum memory, in bits, that any unifilar predictive model of the process (one whose states, like the causal states, are functions of the observed history) must maintain for optimal prediction. It is a non-negative real number and is generally not an integer. A memoryless (IID) process has $\Cmu = 0$, since no past information helps predict the future. A two-state process with equally probable causal states has $\Cmu = 1\bits$; skewed occupations of the same two states give smaller values, with the canonical Even Process \citep[two causal states, $\pi = (2/3, 1/3)$;][]{Crutchfield2012} yielding $\Cmu \approx 0.918\bits$. Higher values correspond to processes with more (or more uniformly occupied) causal states: a uniform distribution over $n$ states gives $\Cmu = \log_2 n\bits$, so $\Cmu$ has no upper bound in principle. A measured $\Cmu \approx 1\bits$ therefore implies at least two distinguishable internal configurations of the process, without by itself fixing their precise number or occupation.

\paragraph{Entropy rate $\hmu$.} The rate of new information production per symbol. The \emachine{} is by construction unifilar and the causal states are a sufficient statistic for the past (knowing the causal state captures everything the history reveals about the future, so no finer detail of the past adds predictive power); under these two properties, and assuming the process is stationary and ergodic, the Shannon entropy rate reduces to the expected single-symbol entropy conditional on the causal state \citep{Crutchfield2012}:
\begin{equation}
    \hmu = -\sum_{s \in \mathcal{S}} \pi(s) \sum_{x \in \mathcal{A}} P(x \mid s) \log_2 P(x \mid s),
    \label{eq:hmu}
\end{equation}
i.e.\ the irreducible randomness that remains even with perfect knowledge of the causal state. Equation~(\ref{eq:hmu}) is therefore an \emachine{}-specific identity; for a general unifilar HMM whose states are not also a sufficient statistic, the right-hand side is only an upper bound on the true $\hmu$.

\paragraph{Excess entropy $E$.} The mutual information between the infinite past and infinite future of the process, $E = I[\overleftarrow{X}; \overrightarrow{X}]$, quantifying the predictive information the past actually contains about the future. $E$ is bounded above by the statistical complexity, $E \leq \Cmu$ \citep{Crutchfield2012}: $E$ is the floor on the memory that any generative model of the process must carry, while $\Cmu$ is the memory the minimal unifilar predictive model (the \emachine) actually carries; the gap between the two is the cost of requiring states that can be read deterministically off the observed history. We do not use $E$ directly in this work, but include it for completeness because the triplet $(\hmu, \Cmu, E)$ is the standard set in the computational-mechanics literature.

\subsection{Relationship to hidden Markov models}
\label{sec:hmm_relation}

Every HMM has a corresponding \emachine, constructed by a deterministic procedure (the mixed-state construction of \citealt{Upper1997}). Relative to the original HMM, the \emachine{} may have more states: the original states are not in general causal states, and resolving them into sufficient statistics for prediction (so that each state is unifilar and makes maximally precise predictions) can require splitting them. Among unifilar HMMs, however, the \emachine{} is minimal: any unifilar HMM of the process has at least as many states. HMMs are themselves a well-established tool in astrophysical time-series analysis; they are used to track continuous gravitational waves from spin-wandering neutron stars \citep{Suvorova2016}, detect pulsar glitches \citep{Melatos2020}, search for gravitational waves from Scorpius~X-1 \citep{Abbott2022ScoX1}, detect continuous nanohertz gravitational waves with pulsar timing arrays \citep{Kimpson2024Kalman, Kimpson2024PulsarTerms, Kimpson2025Background}, suppress instrumental lines in continuous-wave searches \citep{Kimpson2024Mains}, time accreting X-ray pulsars \citep{OLeary2024UKF, OLeary2025RT, OLeary2025Torque}, discover radio pulsars in compact binaries \citep{OLeary2026HMM}, and separate activity states in stellar light curves \citep{Zimmerman2024}. The \emachine{} therefore builds on a modelling framework already familiar in the field, while additionally guaranteeing minimality and a sufficient-statistic interpretation of its states. These guarantees presuppose a stationary, ergodic process, an assumption we revisit for our multi-session sequences in Sections~\ref{sec:cssr_method} and \ref{sec:limitations}.

\subsection{Why computational mechanics matters for FRBs}
\label{sec:why_compmech}

The power spectral density, waiting-time distribution, and scalar complexity measures (approximate entropy, Lyapunov exponents) used in prior FRB analyses are each low-dimensional projections of the full temporal structure. The power spectrum captures second-order correlations but misses higher-order dependencies. The waiting-time distribution describes the marginals but not the sequential dependencies between consecutive intervals. Approximate entropy and Lyapunov exponents provide scalar summaries of regularity and sensitivity to initial conditions but do not identify the hidden states responsible. The \emachine, by contrast, captures the full causal architecture of the process generating the observed temporal pattern. It is the unique minimal sufficient statistic for prediction, and $\Cmu$ complements these scalar summaries by quantifying the minimum predictive memory of the process, a property they do not capture.

\section{Data and Preprocessing}
\label{sec:data}

We analyse three repeating FRB sources with enough bursts for reliable \emachine{} reconstruction. Our synthetic validation (Appendix~\ref{app:validation}) sets a conservative minimum of $N \gtrsim 200$ intra-session waiting times. We treat this as an approximate floor, chosen to keep the reconstruction statistically reliable and to leave a margin for real sequences carrying weaker structure than the clean synthetic benchmark. A fourth well-studied repeater, FRB~20180916B, was excluded because CHIME's transit-mode observing yields only 24 intra-session waiting times, far below the floor. \frbc{}, with 193 intra-session waiting times, sits just under the nominal 200, within the margin the approximate floor is meant to allow; we retain it as a borderline case, and interpret its null result (Section~\ref{sec:emachine_results}) with that caveat. Sources well below this, in the $100 \lesssim N \lesssim 180$ range, could be analysed only with limited statistical power (i.e.\ a reduced probability of detecting genuine temporal structure when it is present), and we leave them to future work. The three sources we analyse are summarised in Table~\ref{tab:sources} and described below. Their burst arrival times come from public archives: the Science Data Bank for the two FAST sources, and the CHIME/FRB catalogue for \frbc{}.

\begin{table*}
    \centering
    \begin{tabular}{@{}lllrrrr@{}}
        \toprule
        Source & Telescope & Reference & $N_\mathrm{burst}$ & $N_\mathrm{session}$ & $N_\mathrm{wt}$ & Median waiting time (s) \\
        \midrule
        \frba & FAST & \citet{Li2021}   & 1652 & 39  & 1613 & 39.5 \\
        \frbb & FAST & \citet{Zhang2022} & 881  & 4   & 877  & 5.1 \\
        \frbc & CHIME & \citet{CHIME2026catalog2} & 353  & 160 & 193  & 149.0 \\
        \bottomrule
    \end{tabular}
    \caption{Summary of the three repeating FRB sources analysed. $N_\mathrm{burst}$ is the total number of detected bursts, $N_\mathrm{session}$ is the number of observing sessions (defined by a 2-hour gap threshold), and $N_\mathrm{wt}$ is the number of intra-session waiting times. The median waiting time is computed from the intra-session intervals only.}
    \label{tab:sources}
\end{table*}

\begin{itemize}
	\item \textbf{FRB~20121102A}. The first known repeater, observed extensively by FAST during a 53-day campaign spanning Modified Julian Date (MJD)~58724--58777 \citep{Li2021}; earlier Arecibo monitoring captured a similar burst-storm episode in 2016 \citep{Hewitt2022}. This dataset comprises 1652 bursts across 39 observing sessions, yielding 1613 intra-session waiting times with a median of 39.5\,s. The waiting times span nearly eight orders of magnitude (from ${\sim}10^{-4}$\,s to ${\sim}10^{4}$\,s), reflecting alternation between periods of intense clustered activity and relative quiescence within sessions.
	\item \textbf{FRB~20201124A}. Observed by FAST during an active episode spanning 2.9\,days (MJD~59482.9--59485.8; \citealt{Xu2022, Zhang2022}). Despite the short observing window, the source's high burst rate yielded 881 bursts across 4 sessions and 877 intra-session waiting times, with the shortest median waiting time (5.1\,s) of the three sources.
	\item \textbf{FRB~20220912A}. A prolific repeater detected by CHIME across a 368-day span (MJD~59833--60201; \citealt{CHIME2026catalog2}). CHIME operates in transit mode, observing each source for approximately 10--15 minutes per transit. This yields 353 bursts across 160 sessions but only 193 intra-session waiting times (median 149\,s), because most transits contain at most two bursts.
\end{itemize}

For each source we compute waiting times only between consecutive bursts within the same observing session, excluding the gaps between sessions, which reflect the telescope's schedule rather than the source's activity. The bursts fall into well-separated clusters, so assigning them to sessions is unambiguous: across all three sources, the largest gap between consecutive bursts within a session is 104 minutes (for \frba; the other two sources are tighter still), while the smallest gap between sessions is 19.1 hours, with no gaps in between. We treat any gap longer than 2 hours as a break between sessions; 2 hours is a round number inside this empty interval, and any threshold between ${\sim}2$ and 19 hours yields the same partition, so no result in this paper depends on the exact value. We adopt each source's published burst list without modification. Whether closely spaced components are catalogued as distinct bursts or as sub-burst microstructure follows the originating catalogue, so the shortest waiting times (down to ${\sim}10^{-4}$\,s for \frba{}) reflect that catalogue's burst definition rather than a criterion imposed here. For each source, the per-session waiting-time sequences are then concatenated end-to-end into a single sequence, which forms the input to the \emachine{} reconstruction (Section~\ref{sec:methods}). This concatenation leaves internal joins between successive sessions; whether the detected memory operates within sessions or across these joins is the central question of the session-resolved analysis in Section~\ref{sec:one_bit} and Appendix~\ref{app:boundary_free}.

\begin{figure}
    \centering
    \includegraphics[width=\columnwidth]{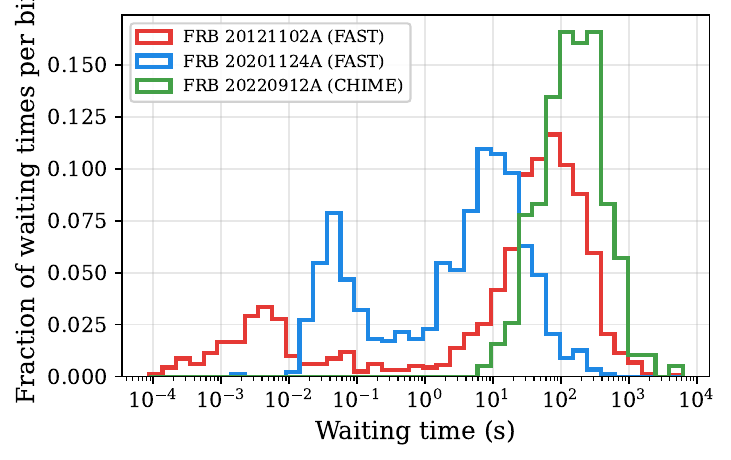}
    \caption{Waiting-time distributions for the three FRB sources, plotted as overlaid step histograms with shared log-spaced bins. The y-axis is the fraction of each source's waiting times falling in each bin, so each histogram sums to unity. The legend identifies each source and its telescope; the number of waiting times and the median waiting time per source are given in Table~\ref{tab:sources}. Both FAST sources span many orders of magnitude in waiting time: \frba{} extends from ${\sim}10^{-4}$\,s to ${\sim}10^{4}$\,s (eight orders) and is bimodal, with distinct populations near ${\sim}10^{-3}$\,s and ${\sim}10^2$\,s, while \frbb{} runs from ${\sim}10^{-3}$\,s to ${\sim}10^3$\,s (six orders). \frbc{} (CHIME) has a much narrower range concentrated around ${\sim}10^2$\,s, set by CHIME's transit-mode cadence.}
    \label{fig:ibi_distributions}
\end{figure}

\begin{figure*}
    \centering
    \includegraphics[width=\textwidth]{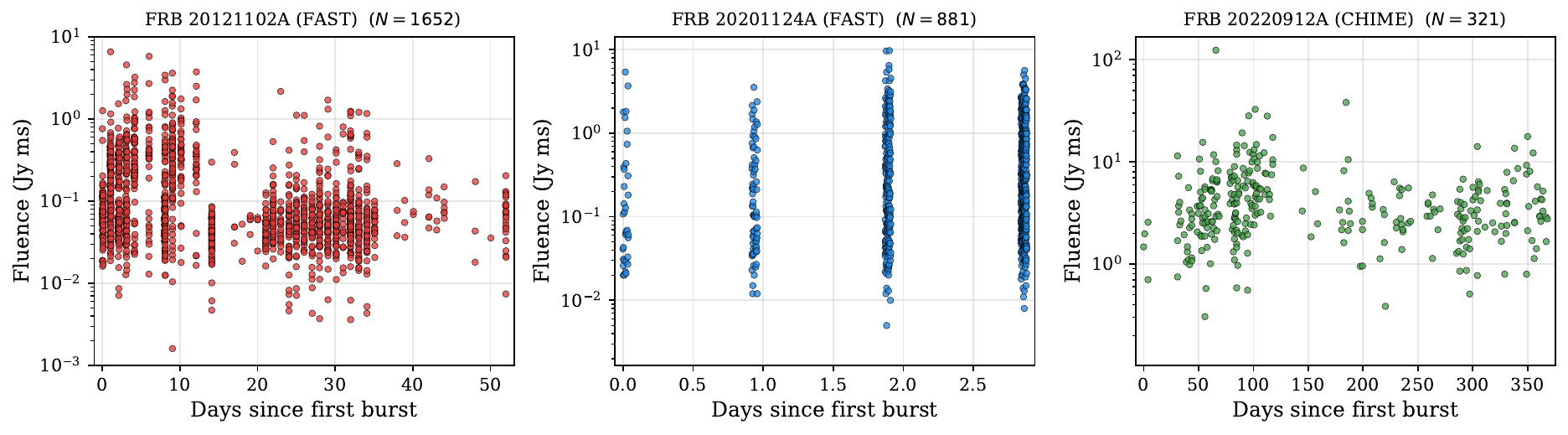}
    \caption{Burst arrival times and fluences for the three FRB sources. Each marker is one detected burst, plotted as fluence (Jy\,ms, log scale) versus time since the first burst in that source's campaign; the x-axis is scaled per panel to the individual campaign duration. The clustering reflects each instrument's observing cadence rather than intrinsic source activity: FAST (\frba{}, \frbb{}) operates in pointed-tracking mode, so bursts are confined to the discrete observing sessions during which the source was being tracked, while CHIME (\frbc{}) is a transit instrument and samples the source for only ${\sim}10$--$15$~min per sidereal day, producing the much sparser scatter spread across the full 368-day campaign. Per-panel titles report the number of plotted bursts; for \frbc{}, 32 of the 353 catalogued bursts have non-finite fluences in the CHIME catalogue and are omitted from this figure (but retained for all timing-based analyses).}
    \label{fig:burst_timeseries}
\end{figure*}

The waiting-time distributions for all three sources are shown in Figure~\ref{fig:ibi_distributions}, while the underlying burst arrival times and fluences for each campaign are displayed in Figure~\ref{fig:burst_timeseries}.

\section{Methods}
\label{sec:methods}

The core analysis pipeline is organised as follows. We first convert each continuous waiting-time sequence into a discrete symbol stream (Section~\ref{sec:symbolisation}) and reconstruct an \emachine{} from it using the CSSR algorithm (Section~\ref{sec:cssr_method}). We quantify finite-sample uncertainty on the resulting $\Cmu$ via block bootstrap (Section~\ref{sec:bootstrap}). Because a finite sample of a memoryless process can yield $\Cmu > 0$ through estimation bias alone, a nonzero estimate does not by itself establish memory; we therefore test the observed $\Cmu$ against surrogates (Section~\ref{sec:surrogates}), which define the memoryless null and distinguish linear from nonlinear memory. A mutual-information diagnostic (Section~\ref{sec:mi_diagnostic}) separates CSSR reconstruction failure from genuine memorylessness. The full pipeline is validated end-to-end on synthetic processes with known causal structure (Section~\ref{sec:pipeline_summary}).

\subsection{Symbolisation}
\label{sec:symbolisation}

CSSR requires a discrete symbol stream as input, and the choice of discretisation determines which temporal features survive into the reconstructed machine. We convert each continuous waiting-time sequence into symbols via quantile discretisation: for a given alphabet size $k$, we sort the waiting times, partition them into $k$ bins of equal occupancy, and replace each waiting time by the index of its bin. This produces a symbol sequence over the alphabet $\mathcal{A} = \{0, 1, \ldots, k-1\}$, where symbol 0 represents the shortest waiting times and symbol $k-1$ the longest. We reconstruct \emachines{} at $k = 2, 3, 4, 5$ and report results across this range as a robustness check on the same underlying temporal structure, not as a set of independent hypotheses.

Equal-frequency (quantile) binning has two advantages over equal-width binning for this application. First, it guarantees that each symbol is well-represented regardless of the shape of the waiting-time distribution, avoiding the empty-bin problem that heavy-tailed distributions create for equal-width bins. Second, the marginal symbol distribution is uniform by construction ($P(X_t = a) = 1/k$ for every $a \in \mathcal{A}$), so any departure from uniformity in the conditional next-symbol distribution $P(X_t \mid \overleftarrow{X})$ is attributable to temporal structure rather than to distributional artefacts.

The discretisation is intentionally lossy: the reconstructed \emachine{} is the \emachine{} of the symbolised process, a coarse-grained projection of the underlying continuous dynamics. The surrogate tests of Section~\ref{sec:surrogates} validate that the excess of $\Cmu$ over the null reflects genuine sequential structure in this projection rather than properties of the binning. \citet{Brodu2022} extend \emachine{} reconstruction to ingest continuous-valued observations directly, with no binning step; applying their method to the FRB waiting-time sequences is a natural extension of this work. An \emachine{} formalism developed specifically for continuous-time, discrete-event processes \citep{Marzen2022} offers a complementary route, since burst arrivals are themselves such a process.

\subsection{CSSR reconstruction}
\label{sec:cssr_method}

We reconstruct \emachines{} from each symbol sequence using the CSSR algorithm \citep{Shalizi2004}, which recovers an \emachine{} from a single observed sequence and makes no parametric assumption about the underlying state structure. We use the implementation in the \texttt{emic} Python package\footnote{\url{https://pypi.org/project/emic/}} \citep{emic}, which has been validated against multiple inference back-ends and exposes all of the standard complexity measures introduced in Section~\ref{sec:measures}. The \texttt{emic} implementation uses a $\chi^2$ homogeneity test for state-splitting; the original Shalizi--Klinkner CSSR \citep{Shalizi2004} defaults to a Kolmogorov--Smirnov variant but lists $\chi^2$ as an equivalent alternative.

CSSR has two hyperparameters: the maximum history length $L$ (the longest past considered when testing whether two histories are predictively equivalent) and the significance level $\alpha$ for the $\chi^2$ state-splitting test (the threshold above which two histories are merged into the same causal state). We adopt $L = 5$ and $\alpha = 0.001$ throughout; the conservative $\alpha$ guards against over-splitting in finite samples, and Appendix~\ref{app:hyperparameters} reports the sensitivity of $\Cmu$ to both choices. Pseudocode for the algorithm is given in Appendix~\ref{app:cssr}.

Increasing the alphabet size $k$ improves temporal resolution and detection power, but spreads a fixed sample over exponentially more candidate histories, so beyond some $k$ the $\chi^2$ state-splitting test loses calibration and the reconstructed $\Cmu$ overshoots. At the opposite extreme, too coarse an alphabet makes the test fail to fire at all, collapsing genuinely structured data to $\Cmu = 0$; this low-$k$ reconstruction failure is diagnosed separately in Section~\ref{sec:mi_diagnostic}. The state-splitting test compares, for each candidate history, the observed counts of the $k$ possible next symbols; these counts form a contingency table with one cell per (history, next-symbol) pair, and the $\chi^2$ approximation to the test statistic is trustworthy only when the expected count in each cell is of order a few or more. With up to $k^L$ candidate histories, each split across its $k$ next-symbol outcomes, the expected count per cell under the null is $\sim N/k^{L+1}$. For our parameters ($L = 5$, $k \in \{2, 3, 4, 5\}$, and $N \in \{193, 877, 1613\}$ for the three sources), this condition holds comfortably at $k = 2$ ($3$--$25$ expected counts per cell) but fails at $k = 4$ ($0.05$--$0.4$) and $k = 5$ ($0.01$--$0.1$), so at these resolutions the test's asymptotic calibration is not formally justified. The conservative $\alpha = 0.001$ partly compensates by demanding strong evidence before splitting, but the $\Cmu$ values at $k \geq 4$ should be interpreted as approximate rather than asymptotically calibrated.

This calibration ceiling is why we do not simply adopt the largest available $k$, even though detection power increases with resolution (Appendix~\ref{app:validation}): the surrogate test of Section~\ref{sec:surrogates}, which calibrates significance empirically against reshuffled data rather than through the $\chi^2$ approximation, controls the false-positive rate at any $k$, but it cannot prevent the over-splitting that inflates the reconstructed $\Cmu$ itself. Our synthetic benchmark (the two-state HMM of Appendix~\ref{app:validation}) has hidden-state entropy $H(\pi) = 0.918\bits$ and a reconstructed \emachine{} complexity of ${\sim}0.96\bits$, slightly above $H(\pi)$ because the \emachine{} of a non-unifilar HMM may resolve additional causal states; at $k = 4$ the median reconstructed $\Cmu$ (${\sim}0.89$--$0.97\bits$ for $N \geq 500$) is consistent with this benchmark scale, whereas at $k = 5$ it rises to ${\sim}1.08\bits$ at the largest $N$, above both reference values. We therefore anchor on $k = 4$ as our fiducial resolution: although its sparse cell counts leave the $\chi^2$ approximation formally unjustified, it is the finest resolution at which the benchmark recovery remains consistent with the generator, with $k = 5$ already overshooting. The benchmark is a proxy for, not a model of, the FRB data, and it informs only this choice of resolution; the detection claims themselves rest on the surrogate comparison of Section~\ref{sec:surrogates}, in which real and surrogate sequences pass through the identical reconstruction, so any residual miscalibration affects both alike. We read the full sweep as a robustness check rather than a menu from which to select the most significant resolution.

CSSR is a consistent estimator (it converges to the true \emachine{} in the long-data limit) only for processes that are stationary with finitely many causal states \citep{Shalizi2004}. If the true process requires infinitely many causal states, CSSR can only capture the structure visible in histories of length $L$ or shorter, and the reconstructed $\Cmu$ then underestimates the true value: it is a lower bound, not a converged estimate. This truncation bias and the finite-sample over-splitting discussed above act in opposite directions, the former deflating the reconstructed $\Cmu$ and the latter inflating it; neither can be corrected exactly, which is why we treat the absolute value of $\Cmu$ as approximate throughout and draw conclusions only from its excess over the surrogate null. The consequences of the stationarity assumption for our concatenated multi-session sequences are taken up in Section~\ref{sec:limitations}.

\subsection{Bootstrap confidence intervals}
\label{sec:bootstrap}

To quantify finite-sample uncertainty on the point estimate $\Cmu$ before testing it against any null, we compute block-bootstrap confidence intervals \citep{Efron1979, Kunsch1989}. For each source and $k$, we draw 200 block-bootstrap replicates of the waiting-time sequence: each replicate is assembled by drawing, with replacement, blocks of consecutive waiting times from the original sequence and concatenating them until the full length $N$ is reached. The block length is selected automatically as the larger of two estimates: the empirical decorrelation length of the waiting-time sequence (the first lag at which the autocorrelation function falls below $1/e$) and the $N^{1/3}$ scaling rule \citep{HallHorowitzJing1995}, rounded to the nearest integer, which acts as a floor when the linear autocorrelation is weak. For all three sources the autocorrelation falls below $1/e$ at the first lag, so the $N^{1/3}$ rule sets the block length: 12, 10, and 6 for \frba, \frbb, and \frbc{} respectively. Because interval widths can be sensitive to the block length for dependent data, we repeated the procedure at half and double these values. The widths change by at most ${\sim}35$ per cent, in the expected direction (longer blocks preserve more serial dependence and give wider intervals), and every interval continues to include zero (Appendix~\ref{app:robustness}). We symbolise each replicate, reconstruct the \emachine, and report the 95\% confidence interval and standard error of $\Cmu$. A naive IID bootstrap would destroy precisely the serial dependence that drives $\Cmu$; the block bootstrap preserves that dependence within blocks while resampling the overall sequence.

\subsection{Surrogate testing}
\label{sec:surrogates}

To test whether the observed $\Cmu$ is consistent with a memoryless process, and to characterise the nature of any detected memory, we generate three families of non-parametric surrogates and compare the real data against each. Each surrogate family answers a different question by destroying a different aspect of the data while preserving others. Each surrogate passes through the same symbolise--reconstruct pipeline as the real data (Section~\ref{sec:symbolisation}), and we report significance as a one-sided $p$-value: the fraction of surrogates with $\Cmu$ at or above the real value. We also report a $z$-score $z = (\Cmu^\mathrm{real} - \langle \Cmu^\mathrm{surr} \rangle) / \sigma_{\Cmu}^\mathrm{surr}$ as an effect-size summary, but because the surrogate $\Cmu$ distributions are non-Gaussian and bounded below by zero, it carries no inferential weight beyond the $p$-value. The three families are described in turn below, followed by the multiple-comparison correction applied across alphabet sizes.

\paragraph{Permutation surrogates.} For each source and each $k$, we generate 1000 surrogate sequences by randomly permuting the waiting-time sequence and re-symbolising with the same quantile boundaries as the original data. The permutation preserves the marginal symbol distribution while destroying all temporal order, providing the primary null for the question ``is the waiting-time sequence memoryless?'' \citep{Theiler1992}. This is the test reported in Section~\ref{sec:emachine_results} as the main result of the paper.

\paragraph{IAAFT surrogates.} To test the stronger null that $\Cmu$ reflects only the linear autocorrelation captured by the autocorrelation function (ACF), we additionally generate 1000 Iteratively Amplitude-Adjusted Fourier Transform (IAAFT) surrogates \citep{SchreiberSchmitz1996} of the raw (continuous) waiting-time sequence. IAAFT preserves both the marginal amplitude distribution and the power spectrum of the source series while destroying higher-order temporal structure; each surrogate is then quantile-symbolised with the same boundaries as the real data and passed through CSSR. Significance against this null indicates that the detected memory has a nonlinear component beyond what the ACF alone captures.

\paragraph{ACF surrogates.} The same 1000 permuted sequences are reused as the null distribution for the autocorrelation function itself: for each source we compute the ACF at lags 1--20 and flag a lag as significant when the real ACF exceeds the 95th percentile of the surrogate ACF distribution. This tests for temporal structure at the level of pairwise correlations, complementing the $\Cmu$-based tests above.

\paragraph{Multiple-comparison correction.} The four tested $k$ values per source are not independent hypotheses but four resolutions of the same underlying structure. We apply a Benjamini--Hochberg false discovery rate (FDR) correction \citep{BenjaminiHochberg1995} per source ($m = 4$ tests each) and report both raw and adjusted $p$-values in Table~\ref{tab:surrogates}.

\subsection{Mutual-information diagnostic}
\label{sec:mi_diagnostic}

CSSR can fail to detect structure that is genuinely present: at coarse alphabets the conditional transition probabilities become nearly uniform, the $\chi^2$ state-splitting test never fires, and the reconstruction returns $\Cmu = 0$ for data that are not memoryless (Section~\ref{sec:cssr_method}). A $\Cmu = 0$ entry is therefore ambiguous between reconstruction failure and genuine memorylessness, two outcomes with qualitatively different physical implications. To distinguish them, we compute the lag-1 mutual information (MI) between consecutive symbols in the symbolised sequence,
\begin{equation}
    I(X_t; X_{t+1}) = \sum_{a, b \in \mathcal{A}} P(a, b) \log_2 \frac{P(a, b)}{P(a)\,P(b)},
    \label{eq:mi}
\end{equation}
where $P(a, b)$ is the empirical joint distribution of consecutive symbol pairs and $P(a)$, $P(b)$ are its marginals, and compare it against the MI of 1000 shuffled copies. If the MI is significantly above the shuffled baseline, the symbolisation has preserved temporal structure in the data even when CSSR returns $\Cmu = 0$. This diagnostic underwrites our interpretation of the $\Cmu = 0$ entries in the main results table, distinguishing those attributable to CSSR coarse-resolution failure from those signalling true memorylessness.

\subsection{Pipeline validation}
\label{sec:pipeline_summary}

Before applying the pipeline to real FRB data, we validate it end-to-end on synthetic processes with known causal structure: the analytic Even and Golden Mean machines, and Poisson and two-state HMM generators. The \texttt{emic} CSSR implementation recovers the analytic machines to within $0.5\%$ of their known $\Cmu$; these machines emit discrete symbols directly, so this test isolates the reconstruction step from symbolisation. The looser benchmark agreement quoted in Section~\ref{sec:cssr_method} for the two-state HMM reflects the full pipeline, including quantile symbolisation of continuous values and a generator whose \emachine{} complexity is itself only bounded, not exactly known. Applied to the synthetic generators, the permutation-surrogate test controls false positives at the nominal rate and recovers the structured (HMM) process with a power that grows with alphabet resolution (${\sim}0.64$--$0.93$ across $k = 4$--$5$ for $N \gtrsim 200$). Full details are given in Appendix~\ref{app:validation}.

\section{Results}
\label{sec:results}

This section reports the temporal-memory analysis for the three sources. Section~\ref{sec:emachine_results} presents the central detection: \emachine{} reconstruction and surrogate testing. Section~\ref{sec:one_bit} then asks where the detected memory lives, separating structure within individual observing sessions from structure in how activity varies across sessions. Section~\ref{sec:robustness} closes with four robustness checks: block-bootstrap confidence intervals, the behaviour of $\Cmu$ across alphabet size, a mutual-information diagnostic that separates genuine memorylessness from CSSR reconstruction failure at coarse resolution, and a fluence-completeness control against session-varying detection sensitivity.

\subsection{Detection of temporal memory}
\label{sec:emachine_results}

\begin{figure*}
    \centering
    \includegraphics[width=\textwidth]{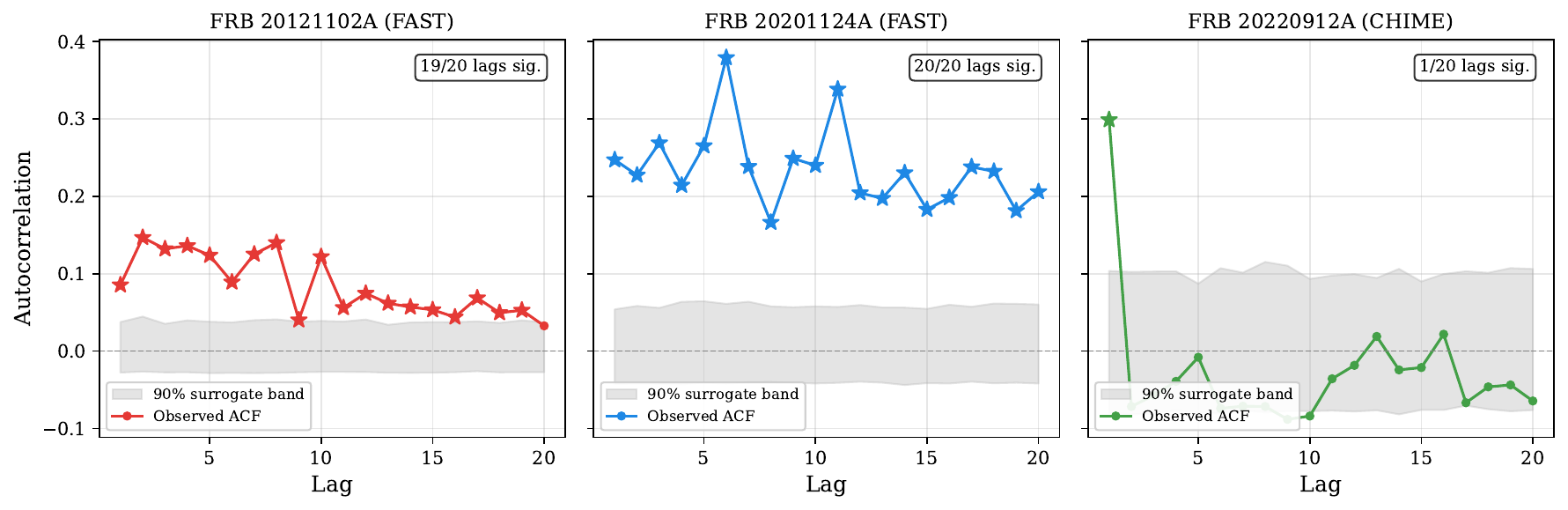}
    \caption{Autocorrelation function of the waiting-time sequence for each source (coloured lines) overlaid with the 5th--95th percentile (90\%) envelope of 1000 permutation surrogates (grey shading). Stars mark lags where the real ACF exceeds the upper one-sided 95th percentile of the surrogate distribution. The number of significant lags (out of 20) is annotated on each panel. Lags count positions in the waiting-time sequence (lag 1 compares consecutive waiting times), not elapsed time.}
    \label{fig:acf}
\end{figure*}

As a model-free preliminary, Figure~\ref{fig:acf} shows the ACF of the waiting-time sequence for each source, overlaid with the 5th--95th percentile (90\%) envelope of the permutation surrogates. \frba{} shows weak but persistent autocorrelation (ACF(1) = 0.085, with 19 of 20 lags exceeding the surrogate 95th percentile) and \frbb{} the strongest sustained autocorrelation (ACF(1) = 0.247, with all 20 lags significant). For \frbc{}, only 1 of 20 lags exceeds the envelope. One exceedance is the rate expected by chance under the null, although this one falls at lag 1, the most diagnostic lag, and its magnitude (ACF(1) $\approx 0.30$, roughly three times the surrogate envelope) is larger than a marginal chance crossing would produce. Short-timescale burst-to-burst correlations have also been reported for this source in independent observations \citep{TotaniTsuzuki2023}, a point taken up in Section~\ref{sec:interpretation}. The mutual-information diagnostic (Section~\ref{sec:robustness}) and the $\Cmu$ tests below are nonetheless null for \frbc{} at every $k$, so we treat the sequence as consistent with memorylessness while noting the lag-1 exceedance. These second-order statistics establish that some temporal structure is present in the FAST sequences, but they cannot say how much predictive memory that structure carries, nor what minimal model generates it; quantifying both is the task of the \emachine{} reconstruction.

\begin{figure*}
    \centering
    \includegraphics[width=\textwidth]{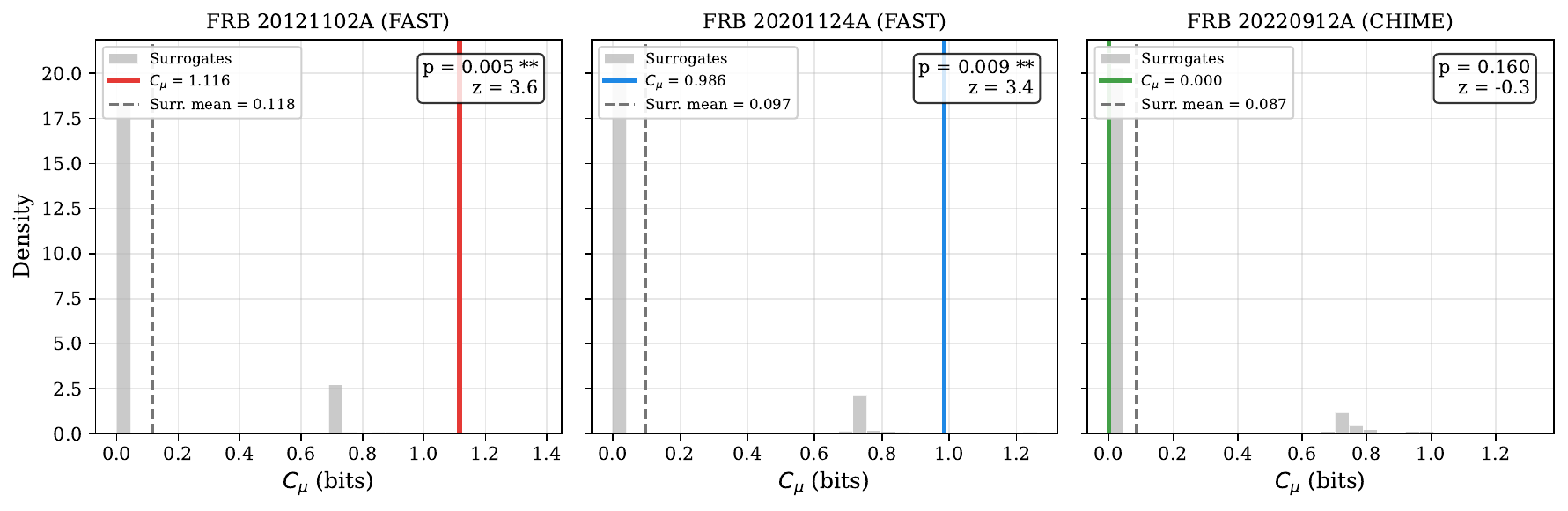}
    \caption{Surrogate comparison at $k = 4$. Each panel shows the distribution of $\Cmu$ from 1000 permutation surrogates (grey histogram) and the real $\Cmu$ (vertical coloured line). Dashed grey lines mark the surrogate mean. The $p$-value and $z$-score are annotated; double asterisks mark $p < 0.01$. Both FAST sources are significant ($p < 0.01$); \frbc{} is not. The surrogate distribution is bimodal (a pile-up at $\Cmu \approx 0$ with a secondary cluster near 0.7--0.8\,bits); this reflects CSSR's binary state-splitting behaviour at finite $N$ rather than physical bimodality, and is discussed further in Appendix~\ref{app:robustness}.}
    \label{fig:surrogates}
\end{figure*}

Figure~\ref{fig:surrogates} compares the real $\Cmu$ for each source at $k = 4$ against the distribution of $\Cmu$ from 1000 permutation surrogates (Section~\ref{sec:surrogates}); Table~\ref{tab:surrogates} reports the full $k$-sweep. Both FAST sources show $\Cmu$ significantly above surrogates at $k \geq 4$. \frba{} yields $\Cmu = 1.116\bits$ at $k = 4$ ($p = 0.005$, $z = 3.6$) and $\Cmu = 0.831\bits$ at $k = 5$ ($p = 0.028$, $z = 2.4$). \frbb{} yields $\Cmu = 0.986\bits$ at $k = 4$ ($p = 0.009$, $z = 3.4$) and $\Cmu = 0.900\bits$ at $k = 5$ ($p = 0.014$, $z = 2.9$). For both sources the surrogate distribution has mean $\Cmu \approx 0.1\bits$, with the real values lying well above its bulk ($p < 0.05$ in all four tests, and $p \leq 0.01$ for both $k = 4$ detections). \frba{} also reaches nominal significance at $k = 2$ ($\Cmu = 0.918\bits$, $p = 0.039$); this detection does not survive the multiple-comparison correction below ($p_\mathrm{adj} = 0.052$), and we do not count it among the robust detections. These canonical values are reconstructed from the full concatenated multi-session sequence; Section~\ref{sec:one_bit} and Appendix~\ref{app:boundary_free} establish that for both FAST sources this $\Cmu$ reflects predictive structure on inter-session timescales rather than within-session memory.

\frbc{} shows $\Cmu = 0$ at all four $k$ values. At $k = 4$, where the FAST sources show the strongest signal, \frbc{} has $p = 0.160$, not significant. At $k = 2, 3, 5$, the $p$-values are 1.000. The \emachine{} for \frbc{} is trivial (a single state with uniform emissions) at every resolution tested. As shown in Section~\ref{sec:bias_results}, this null result may reflect CHIME's short transit-mode observing windows rather than intrinsic memorylessness.

Testing four alphabet sizes across three sources gives 12 hypothesis tests, so Table~\ref{tab:surrogates} also reports Benjamini--Hochberg \citep{BenjaminiHochberg1995} FDR-adjusted $p$-values. The correction can be applied in two ways: within each source, treating its four $k$ values as one family ($m = 4$ tests), which asks whether that source individually shows memory at any resolution; or pooling all 12 tests ($m = 12$), which is stricter because it guards against any source--resolution combination in the study reaching significance by chance. The $k = 4$ detections survive the per-source correction at $p_\mathrm{adj} = 0.020$ (\frba) and $p_\mathrm{adj} = 0.028$ (\frbb, whose $k = 5$ result also survives at $0.028$); under the stricter pooled correction the strongest detections sit at $p_\mathrm{adj} = 0.054$, just above the conventional threshold. Our primary claim rests on the per-source correction, with the pooled numbers reported as a more conservative sensitivity check; the rationale for each choice is discussed alongside the other sensitivity analyses in Appendix~\ref{app:hyperparameters}.

\begin{table*}
    \centering
    \begin{tabular}{@{}llcccc@{}}
        \toprule
        Source & $k$ & $\Cmu$ (bits) & $p_\mathrm{raw}$ & $p_\mathrm{adj}^{m=4}$ & $p_\mathrm{adj}^{m=12}$ \\
        \midrule
        \frba & 2 & 0.918 & \textbf{0.039} & 0.052 & 0.094 \\
              & 3 & 0.000 & 1.000 & 1.000 & 1.000 \\
              & 4 & 1.116 & \textbf{0.005} & \textbf{0.020} & 0.054 \\
              & 5 & 0.831 & \textbf{0.028} & 0.052 & 0.084 \\
        \addlinespace
        \frbb & 2 & 0.000 & 0.084 & 0.112 & 0.168 \\
              & 3 & 0.000 & 0.112 & 0.112 & 0.192 \\
              & 4 & 0.986 & \textbf{0.009} & \textbf{0.028} & 0.054 \\
              & 5 & 0.900 & \textbf{0.014} & \textbf{0.028} & 0.056 \\
        \addlinespace
        \frbc & 2 & 0.000 & 1.000 & 1.000 & 1.000 \\
              & 3 & 0.000 & 1.000 & 1.000 & 1.000 \\
              & 4 & 0.000 & 0.160 & 0.640 & 0.240 \\
              & 5 & 0.000 & 1.000 & 1.000 & 1.000 \\
        \bottomrule
    \end{tabular}
    \caption{Permutation-surrogate tests of the statistical complexity $\Cmu$ across all sources and alphabet sizes $k$. $p_\mathrm{raw}$ is the fraction of 1000 surrogates with $\Cmu \geq \Cmu^\mathrm{real}$; $p_\mathrm{adj}^{m=4}$ and $p_\mathrm{adj}^{m=12}$ are Benjamini--Hochberg FDR-adjusted values under the per-source ($m = 4$ tests per source) and pooled ($m = 12$ tests) corrections. Bold marks $p < 0.05$ in each $p$ column. $\Cmu$ is rounded to three decimal places; entries displayed as $0.000$ can be non-zero at machine precision (${\lesssim}10^{-8}\bits$), and their $p$-values, computed against the unrounded value, can therefore be below unity. Block-bootstrap 95\% confidence intervals on the $\Cmu$ point estimates are reported separately in Table~\ref{tab:bootstrap}.}
    \label{tab:surrogates}
\end{table*}

\begin{figure*}
    \centering
    \resizebox{\textwidth}{!}{%
    \begin{tikzpicture}[
        >=Stealth, semithick,
        every state/.style={
            fill=blue!7, draw=black, line width=0.9pt,
            minimum size=1.35cm, align=center, inner sep=0pt,
        },
        el/.style={
            align=center, font=\footnotesize, inner sep=2pt,
            fill=white, fill opacity=0.9, text opacity=1, rounded corners=1pt,
        },
        every loop/.style={looseness=5},
        title/.style={font=\bfseries, align=center},
    ]
    \node[title] at (0,4.8) {\frba{}\\[2pt]3 states, $\Cmu{=}1.116$ bits};
    \node[state] (a0) at ( 0.0, 2.4) {\large$s_1$\\[-1pt]\scriptsize$\pi{=}0.70$};
    \node[state] (a1) at (-2.4,-1.0) {\large$s_2$\\[-1pt]\scriptsize$\pi{=}0.23$};
    \node[state] (a2) at ( 2.4,-1.0) {\large$s_3$\\[-1pt]\scriptsize$\pi{=}0.07$};
    \path (a0) edge[loop above]
          node[el,above] {\tl{0}{0.24}\;\;\tl{1}{0.22}\;\;\tl{3}{0.29}} (a0);
    \path (a1) edge[bend left=12]                 
          node[el,left] {\tl{0}{0.23}\\\tl{2}{0.27}\\\tl{3}{0.18}} (a0);
    \path (a0) edge[bend left=12]                 
          node[el,pos=0.5,right] {\tl{2}{0.25}} (a1);
    \path (a2) edge[bend right=10]                
          node[el,right] {\tl{2}{0.15}\\\tl{3}{0.10}} (a0);
    \path (a1) edge[bend left=10]                 
          node[el,above] {\tl{1}{0.32}} (a2);
    \path (a2) edge[bend left=10]                 
          node[el,below] {\tl{0}{0.41}\\\tl{1}{0.33}} (a1);
    \node[title] at (8.0,4.8) {\frbb{}\\[2pt]2 states, $\Cmu{=}0.986$ bits};
    \node[state] (b0) at (6.7,0.5) {\large$s_1$\\[-1pt]\scriptsize$\pi{=}0.57$};
    \node[state] (b1) at (9.7,0.5) {\large$s_2$\\[-1pt]\scriptsize$\pi{=}0.43$};
    \path (b0) edge[loop left]
          node[el,left] {\tl{0}{0.24}\\\tl{1}{0.32}\\\tl{2}{0.27}} (b0);
    \path (b1) edge[loop right]
          node[el,right] {\tl{0}{0.26}\\\tl{1}{0.19}\\\tl{3}{0.33}} (b1);
    \path (b0) edge[bend left=22] node[el,above] {\tl{3}{0.17}} (b1);
    \path (b1) edge[bend left=22] node[el,below] {\tl{2}{0.22}} (b0);
    \node[title] at (14.6,4.8) {\frbc{}\\[2pt]1 state, $\Cmu{=}0.000$ bits};
    \node[state] (c0) at (14.6,0.5) {\large$s_1$\\[-1pt]\scriptsize$\pi{=}1.00$};
    \path (c0) edge[loop above]
          node[el,above] {\tl{0}{0.25}\;\;\tl{1}{0.26}\;\;\tl{2}{0.24}\;\;\tl{3}{0.25}} (c0);
    \end{tikzpicture}%
    }
    \caption{Reconstructed \emachines{} at $k = 4$ for all three sources, shown in left-to-right order: \frba{}, \frbb{}, \frbc{}. Nodes represent causal states, labelled with their stationary probability $\pi$; directed edges represent transitions, labelled in the format \emph{symbol:probability}, giving the emitted symbol and its conditional emission probability from the source state. The symbol index is colour-coded (\textcolor{sym0}{$0$}, \textcolor{sym1}{$1$}, \textcolor{sym2}{$2$}, \textcolor{sym3}{$3$}) to set it apart from the probability value, and a single edge may bundle several symbols where distinct emissions from one state lead to the same successor. Zero-probability states (CSSR over-splitting artefacts) are omitted. Emission probabilities are rounded to two decimal places, so the outgoing probabilities from a state may not sum exactly to one (e.g.\ $s_3$ of \frba{} sums to 0.99). Both FAST sources yield multi-state machines with $\Cmu \approx 1\bits$, while \frbc{} yields a trivial single-state machine ($\Cmu = 0$), consistent with memoryless emission.}
    \label{fig:epsilon_machines}
\end{figure*}
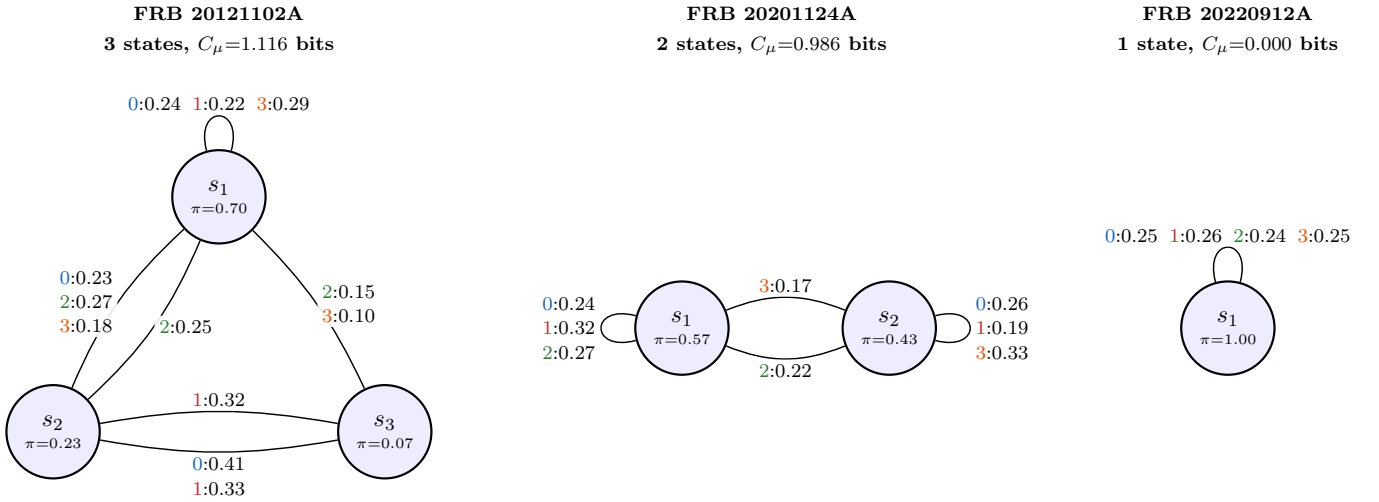

The reconstructed \emachines{} at $k = 4$ are shown in Figure~\ref{fig:epsilon_machines}. \frba{} yields a three-active-state machine, with the dominant state ($\pi = 0.70$) emitting all four symbols at near-equal probability and two minor states ($\pi = 0.23$ and $\pi = 0.07$) with distinct emission profiles. \frbb{} yields a two-state machine with an asymmetric stationary distribution (dominant state $\pi = 0.57$, minor state $\pi = 0.43$), where each state preferentially emits different symbols. \frbc{} yields a trivial single-state machine with uniform emissions, the hallmark of an IID process. The corresponding entropy rates (Equation~\ref{eq:hmu}) at $k = 4$ are $\hmu = 1.97\bitssym$ (bits per symbol) for both FAST sources and effectively the IID value $\log_2 4 = 2\bitssym$ for \frbc{}'s single-state machine; the small deficits from that value for the FAST sources reflect the modest predictive structure that the surrogate tests establish.

To test whether the detected memory extends beyond linear autocorrelation, we additionally generate IAAFT surrogates \citep{SchreiberSchmitz1996}, which preserve the marginal distribution and power spectrum (and hence the linear ACF) while destroying higher-order structure. \frba{} marginally exceeds the IAAFT distribution at $k = 4$ ($p = 0.032$, $z = 2.3$), suggesting a nonlinear component to its memory, while \frbb{} does not ($p \geq 0.098$ at all $k$), consistent with its strong linear autocorrelation accounting for most of the detected $\Cmu$ (though non-significance against IAAFT does not exclude a smaller nonlinear contribution). Full results are tabulated in Appendix~\ref{app:iaaft}.

\subsection{Where does the memory live?}
\label{sec:one_bit}

A non-zero $\Cmu$ on the concatenated waiting-time sequence could reflect within-session causal structure (the per-burst emission mechanism remembering past bursts), inter-session activity-rate variation (the source occupying different bursting regimes across sessions), or some combination of the two. The sequence concatenates many separate observing sessions end to end, so it contains internal session boundaries, the joins between consecutive sessions. Within-session memory lies inside a session; inter-session structure lives across these joins. To disentangle the two, we run four analyses at $k = 4$ with $L = 5$. Three are \emph{session-boundary controls}, which perturb or restrict the data and test whether the signal survives: (i) a session-shuffle test (100 permutations of session order, with within-session waiting times intact); (ii) per-session quantile binning (bin edges recomputed within each session, requiring $\geq 30$~waiting times per session for inclusion); and (iii) a reconstruction restricted to each source's single longest session. The fourth, (iv), is a \emph{boundary-free reconstruction}, which removes cross-session histories by construction: CSSR's suffix tree is built per session, so no history of length $L$ spans a session boundary. The boundary-free analysis is developed in full in Appendix~\ref{app:boundary_free}, which carries most of the quantitative detail behind this separation.

\paragraph{(i) Session-shuffle test.} For \frba, only one of the 100 session-order permutations reaches the observed $\Cmu$, giving $p = 0.020$ under the Monte-Carlo $(r+1)/(n+1)$ estimate \citep{PhipsonSmyth2010}: session ordering matters, and the signal is not merely an artefact of combining heterogeneous sessions in an arbitrary order. With 100 permutations this sits near the resolution floor of the test, so the precise value should not be over-read; the qualitative conclusion that ordering carries signal does not depend on it. For \frbb, $p = 0.376$: the signal does not depend on session ordering, consistent with the memory arising from the contrast between sessions rather than from structure within them.

\paragraph{(ii) Per-session quantile binning.} Under this scheme, \frba{} retains $\Cmu = 0.955\bits$ while \frbb{} drops to $\Cmu = 0.000\bits$. This test removes inter-session distributional differences but still operates on a concatenated symbol stream, so it does not by itself eliminate cross-session histories in the CSSR suffix tree.

\paragraph{(iii) Longest-session analysis.} For \frba, the longest session (121~waiting times) yields $\Cmu = 0.598\bits$, suggestive of within-session structure but on a sample below our $N \approx 200$ minimum-sample requirement (Appendix~\ref{app:validation}). For \frbb, the longest session (541~waiting times, well above this requirement) yields $\Cmu = 0.000\bits$: no within-session structure detected.

\paragraph{(iv) Boundary-free reconstruction.} For \frba, the boundary-free $\Cmu$ at $k = 4$ drops to zero: the canonical $\Cmu = 1.116\bits$ is driven predominantly by cross-session histories. For \frbb, the boundary-free $\Cmu = 0.985\bits$ is essentially unchanged from the canonical 0.986\,bits, as expected since \frbb{} has only three internal session boundaries (${\sim}1.7\%$ of histories) which cannot carry the signal on their own. Tested against a within-session-shuffle null that preserves per-session waiting-time multisets, neither source shows significant within-session predictive memory at any $k$ after multiple-comparison correction; \frba{} retains borderline evidence at $k = 5$ ($p_\mathrm{raw} = 0.033$, $p_\mathrm{adj} = 0.134$ across the $k$-sweep), echoing the longest-session and IAAFT results (Appendix~\ref{app:boundary_free}).

These analyses converge on a common picture: for both FAST sources, the detected $\Cmu \approx 1\bits$ reflects predictive structure operating on inter-session timescales. Each source apparently switches between activity-rate regimes; for \frba{} the sequence of those regimes across sessions itself carries information, while for \frbb{} the information lies in the contrast between heterogeneous sessions. The two sources differ in the timescale of this switching: \frbb{}'s four sessions span 2.9 days, so the regime evolution it samples is on the hours-to-day scale; \frba{}'s 39 sessions span 53 days, so its regime evolution can encompass hours to weeks. \frba{} additionally shows weak hints of within-session structure (the $\Cmu = 0.598\bits$ longest-session result and borderline boundary-free $\Cmu$ and lag-1 MI signals at the finest alphabet resolution), but none of these survive the threshold applied throughout this paper ($p < 0.05$ after Benjamini--Hochberg correction), and confirming within-session memory would require longer continuous sessions than are currently available. For \frbb{}, in contrast, the well-powered single-session reconstruction detects no within-session structure. Applying CSSR to the concatenated non-stationary sequence operates outside the formal stationarity guarantee of the \emachine{} theorem \citep{Shalizi2001}, so the multi-session $\Cmu$ for both sources should be interpreted as quantifying inter-session regime structure rather than per-burst predictive memory.

The within-session-shuffle null adds a final piece of corroboration: for \frbb{}, surrogates with all within-session ordering destroyed still reproduce $\Cmu \sim 0.45\bits$ from the session-level marginal contrasts alone, so a substantial part of its canonical signal requires no ordering information whatsoever (Table~\ref{tab:boundary_free_surr}). What these inter-session regimes are physically, and what they do and do not imply about the source, is taken up in Section~\ref{sec:interpretation}.

\subsection{Robustness checks}
\label{sec:robustness}

Three further checks probe the statistical standing of the detection; the quantitative detail behind each is collected in Appendix~\ref{app:robustness}.

First, block-bootstrap 95\% confidence intervals on the $\Cmu$ point estimates (Section~\ref{sec:bootstrap}; Table~\ref{tab:bootstrap}) are wide, and all include zero. This reflects the bimodal character of the bootstrap distribution (individual replicates yield either $\Cmu \approx 0$ or $\Cmu \approx 1$, the binary firing behaviour of CSSR's $\chi^2$ state-splitting test at finite $N$) rather than consistency with memorylessness; the same collapse occurs on known-structured synthetic data (Appendix~\ref{app:validation}). The bootstrap quantifies how imprecisely the absolute value of $\Cmu$ is determined; the rejection of memorylessness rests on the surrogate test, which asks a different question and is unaffected.

Second, the ranking of the three FRB sources is robust across alphabet size (Figure~\ref{fig:cmu_vs_k}): \frbc{} yields $\Cmu = 0$ at every resolution, while the FAST sources exceed the surrogate band wherever CSSR detects structure. At $k = 2$ and $3$ the FAST sources collapse to $\Cmu = 0$ for some configurations; this is a CSSR reconstruction artefact at coarse resolution, not an absence of structure.

Third, the mutual-information diagnostic (Section~\ref{sec:mi_diagnostic}; Figure~\ref{fig:mi}) confirms that the coarse-resolution collapses are reconstruction artefacts: for both FAST sources the lag-1 MI of the symbolised sequence is significantly above the shuffled baseline at every tested $k$ (at $k = 2$, \frba{} $p = 0.032$ and \frbb{} $p = 0.026$), so the coarse-resolution $\Cmu = 0$ entries reflect CSSR failing to fire where temporal structure is demonstrably present in the symbols. For \frbc{} the MI is not significant at any $k$ ($p > 0.1$): the symbol sequence is indistinguishable from IID at every resolution, consistent with its $\Cmu = 0$.

A fourth control addresses instrumental rather than statistical robustness: because the detected memory is inter-session, session-varying detection sensitivity could in principle mimic it (Section~\ref{sec:limitations}). Appendix~\ref{app:fluence_control} repeats the detection on subsamples restricted to bursts bright enough to be detected regardless of session-to-session sensitivity (above the fluence completeness threshold). To check whether any loss of signal under this cut reflects the smaller sample rather than the removal of faint bursts, each cut is compared against subsamples of the same size with bursts removed at random. \frbb{}'s detection survives cuts up to twice the published completeness threshold, so its signal is not carried by bursts near the detection limit; for \frba{} the control is power-limited: the cut destroys the detection, but removing the same number of bursts at random usually destroys it too, so the loss reflects the smaller sample rather than the removal of faint bursts specifically.

\section{Discussion}
\label{sec:discussion}

The preceding section established that two of the three sources carry temporal memory of order one bit, that this memory operates on inter-session rather than within-session timescales, and that the third source's null result is ambiguous. This section interprets those findings. We develop the physical interpretation (Section~\ref{sec:interpretation}), test whether the regime sequence carries more structure than a two-state rate-switching process can reproduce (Section~\ref{sec:literature}), draw out the implications for observing strategy and the status of \frbc{} (Section~\ref{sec:bias_results}), and set out what waiting-time data alone can and cannot establish (Section~\ref{sec:limitations}).

\subsection{Interpreting the detected memory}
\label{sec:interpretation}

For both FAST sources $\Cmu$ lies well above the bulk of the permutation-surrogate distribution ($p \leq 0.01$): the waiting-time sequence is not memoryless. Its absolute value is less precisely constrained, varying between 0.83 and 1.12\,bits depending on the alphabet size $k$, with wide bootstrap confidence intervals (Section~\ref{sec:robustness}). The measured values are consistently of order one bit, which in the language of computational mechanics means that the process requires at least two distinguishable hidden modes for optimal prediction (since $\Cmu = 1\bits$ is the entropy of a fair binary variable; more states with asymmetric weights could also produce $\Cmu \approx 1$). As established in Section~\ref{sec:one_bit}, these hidden modes are best understood as the activity-rate regimes that each source occupies across its observing campaign, with the timescale of regime switching distinguishing the two sources.

Both FAST sources yield $\Cmu$ of similar magnitude despite different autocorrelation structures and different reconstructed machine topologies (\frba: three active states; \frbb: two active states). \frba{} shows weak but persistent autocorrelation (ACF(1) = 0.085, 19/20 lags significant), while \frbb{} shows strong persistent autocorrelation (ACF(1) = 0.247, 20/20 lags significant). We cannot determine from two sources alone whether the similar $\Cmu$ values reflect a shared underlying constraint on the burst-generating process or are coincidental given the uncertainties. The two sources require a comparable amount of predictive memory despite their different correlation structures; the nature of that dependence, which we examine below, differs even where the amount does not.

The localisation results of Section~\ref{sec:one_bit} sharpen the distinction between the two sources. For \frba{}, the session ordering itself is predictive: rearranging the sessions destroys the signal. For \frbb{}, every ordering-sensitive control is null, and what remains is session-level rate heterogeneity: four sessions with distinguishably different waiting-time distributions. \frbb{}'s $\Cmu$ is therefore best read as the memory required to track which kind of session the source is in, not as evidence for an ordered regime sequence. The binning sensitivity reinforces this reading (Appendix~\ref{app:hyperparameters}): \frbb{}'s detection requires quantile discretisation, which by construction encodes inter-session distributional contrast, and vanishes under log-spaced binning, whereas \frba{}'s detection survives the change of scheme.

These results point to a simple picture: both sources are well described by a process that alternates between a small number of activity-rate regimes, with individual sessions approximately stationary within a given regime and the regime varying between sessions. The boundary-free analysis explains why \frba's full-concatenation $\Cmu = 1.116\bits$ exceeds the per-session quantile result ($\Cmu = 0.955\bits$) and the longest-session result ($\Cmu = 0.598\bits$): the larger value picks up cross-session histories that encode the sequence of activity regimes, while the smaller values progressively remove that contribution. Many physical mechanisms are consistent with this picture. One class is magnetospheric models in which the source's activity is set by a slowly evolving global configuration (a twisted magnetosphere whose state changes on timescales longer than an individual observing session; \citealt{Wadiasingh2019, Beloborodov2020, Lu2020}) whose sequence of occupied configurations is what the \emachine{} resolves. Other classes include propagation effects in a structured local environment and intermittent activity in a companion-perturbed magnetosphere. The waiting-time analysis does not by itself distinguish among these alternatives.

The detection of $\Cmu$ significantly above zero is independent of any parametric model of the waiting-time distribution (though not of the discretisation scheme; Appendix~\ref{app:hyperparameters}): the waiting-time process has temporal memory, and any viable physical model must reproduce this. The localisation gives the constraint its second side: the memory lives between sessions, and within sessions the timing is consistent with memoryless emission at current sensitivity, so models that generically predict strong burst-to-burst correlation on second-to-minute scales receive no support from these data at the resolutions our symbolisation probes.

This within-session statement must, however, be reconciled with \citet{TotaniTsuzuki2023}, who report aftershock-like burst-to-burst correlations in these same three sources. Their signal is strongest at the shortest separations, where quantile symbolisation compresses the entire short-waiting-time population into a single symbol, so a rate enhancement largely confined within one bin's dynamic range is mostly invisible to our test; the borderline within-session signals at our finest resolution (Appendix~\ref{app:boundary_free}) may be a trace of it. The two results are therefore not directly contradictory, but a quantitative reconciliation, ideally under a common burst definition, is left to future work. Taken together with the inter-session localisation of Section~\ref{sec:one_bit}, this leaves a two-sided constraint on physical models: any viable model must produce inter-session activity-rate regimes, with an ordering that is itself informative in the case of \frba{}, while remaining consistent with memoryless within-session timing at the resolutions probed here. This constraint also bears on prior claims of memory in these same sources: \citet{WangWuDai2023} and \citet{SangLin2024} report Hurst-exponent persistence on minute-to-hour (within-session) timescales, precisely the scales where our boundary-free and within-session-shuffle controls find no surviving predictive structure (though the test is power-limited for \frba; Section~\ref{sec:limitations}). Persistence statistics and predictive complexity are different objects; Appendix~\ref{app:literature} discusses how our result relates to these prior claims.

The IAAFT surrogate analysis (Appendix~\ref{app:iaaft}) provides further insight into the nature of this memory at the multi-session level. For \frba, there is marginal evidence ($p = 0.032$ at $k = 4$) that the temporal structure of the concatenated sequence includes nonlinear components beyond the autocorrelation function. For \frbb, the IAAFT surrogates produce $\Cmu$ distributions consistent with the observed value, suggesting that the strong linear ACF (ACF(1) = 0.247) may account for most of the detected $\Cmu$. In both cases the inter-session process requires $\sim$1\,bit of predictive memory, but the dependence structure differs: \frba{} appears to have a nonlinear component in how its activity regimes sequence between sessions, while \frbb{}'s sequence is consistent with predominantly linear dependence.

The \emachine{} framework provides more than a scalar detection of temporal memory. Consider \frbb's reconstructed machine at $k = 4$: a two-state model with stationary occupations $\{0.57, 0.43\}$, where each state emits different symbol distributions. This is a minimal generative model: it specifies the states, their occupancy probabilities, and the transition structure. The ACF detects that ``consecutive waiting times are correlated''; the \emachine{} tells us \emph{how} they are correlated, via a model that can be simulated, tested against future data, and compared across sources (though the reconstructed state topology is less reliable than $\Cmu$ itself; Section~\ref{sec:limitations}). $\Cmu$ also quantifies the memory: $\sim$1\,bit means optimal prediction of the observed sequence requires at least two distinguishable states, one bit's worth of information about which activity regime is active over an observing session. This is a tighter constraint than ``waiting times are autocorrelated''.

Two superficially similar claims must be kept apart. We establish that the observed multi-session waiting-time sequence requires at least two activity-rate regimes for optimal prediction; we do not establish that the source itself possesses two or more internal physical configurations. A single-state engine with a continuously drifting mean rate, sampled at sessions that cluster into a few rate levels, would yield the same $\Cmu$, as would an external modulator (orbital, propagation, environmental) imposing rate variation on a structurally simpler source. The \emachine{} result constrains the predictive structure of the observed sequence, not the dimensionality of the source's internal state space, and we do not identify the regimes with specific physical mechanisms. This degeneracy is not specific to computational mechanics: any inference from burst arrival times alone (parametric waiting-time fits, persistence statistics, periodicity searches) cannot separate an intrinsic state sequence from externally imposed rate variation. What the analysis does constrain is the timescale and magnitude of variability: whatever modulates FRB emission, on inter-session timescales it must vary by enough to carry at least one bit of predictive information about future emission rate. This rules out memoryless (constant-rate Poisson) emission and quantifies the ordered rate variation that any viable model (whether a stationary hidden-state process or a genuinely non-stationary one) must accommodate.

\subsection{Comparison with a Markov-modulated Poisson process}
\label{sec:literature}

Could the detected memory be merely the signature of a rate-switching process? Non-Poissonian, clustered waiting-time models have been fitted to these sources: a Weibull renewal process \citep{Oppermann2018, Cruces2021} and a time-varying-rate Poisson process \citep{Jahns2023}. The natural rate-switching generalisation of these models is a Markov-modulated Poisson process (MMPP). In the MMPP, a hidden state $\zeta_n \in \{1, 2\}$ governs each waiting time, and transitions between fast and slow rates carry predictive information about the next waiting time's expected length. We test this directly.

We implemented the MMPP as a discrete-time hidden Markov model: transitions follow a $2 \times 2$ stochastic matrix $\mathbf{T}$ and each waiting time is drawn from $\mathrm{Exp}(\lambda_{\zeta_n})$. For each source we fitted the MMPP by Expectation--Maximisation (four free parameters), drew 1000 synthetic waiting-time sequences of length matched to the observed $N$, symbolised each at $k = 4$ using its own quantile boundaries, reconstructed an \emachine{} for each, and recorded the resulting $\Cmu$. These baselines condition on the best-fitting point estimate; uncertainty in the fitted MMPP parameters is not propagated into the synthetic $\Cmu$ distributions. Symbolising each draw with its own quantile boundaries rather than those of the real data matches the procedure applied to the observations, where the quantiles are estimated from the sequence in hand. It also prevents the real marginal from leaking into the synthetic comparison, so the test isolates sequential structure rather than marginal differences. One structural mismatch remains: the synthetic draws are gap-free sequences of matched length, whereas the real data concatenate sessions separated by long unobserved gaps across which any hidden state continues to evolve; part of the $\Cmu$ excess over the fitted MMPP could therefore reflect this sampling mismatch rather than dynamics beyond rate-switching, a caveat that the higher-order structured nulls discussed below would also address. For comparison with the published parametric fits \citep{Cruces2021, Oppermann2018}, we additionally fitted a secondary IID two-component exponential mixture (three free parameters, fitted by Expectation--Maximisation, likewise 1000 synthetic draws); this null shares the marginal waiting-time distribution of a corresponding MMPP but cannot reproduce $\Cmu > 0$ from temporal ordering by construction.

\begin{figure*}
    \centering
    \includegraphics[width=\textwidth]{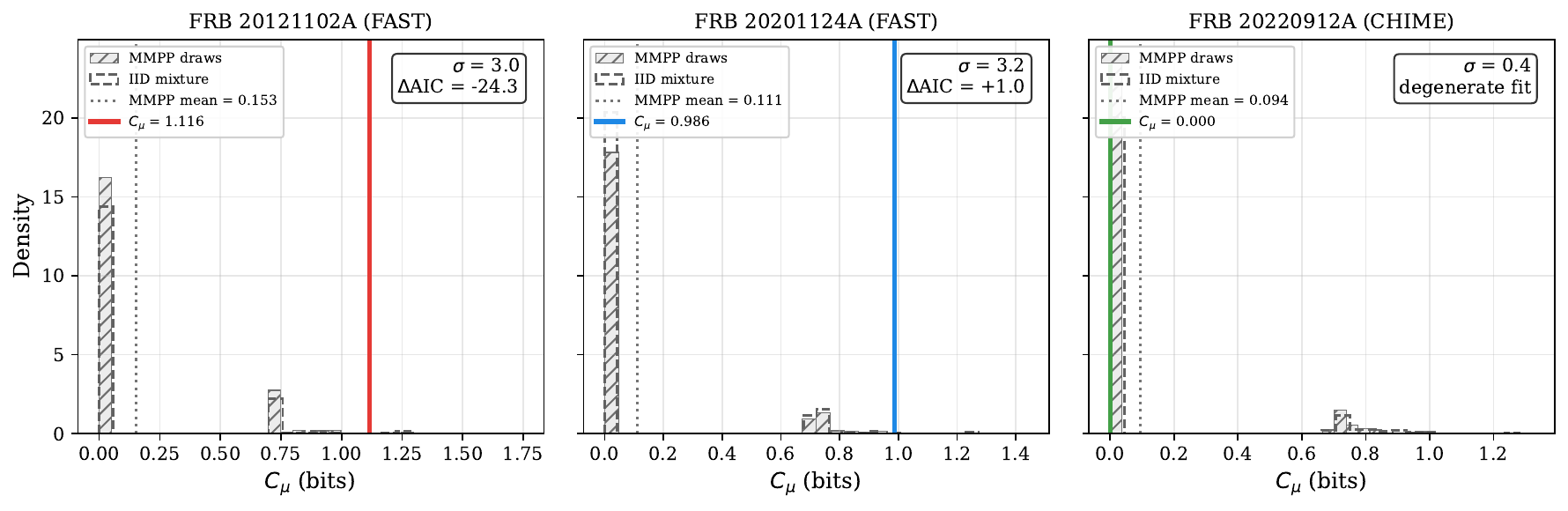}
    \caption{Observed statistical complexity $\Cmu$ (coloured vertical line) against the distribution of $\Cmu$ from 1000 synthetic draws of the fitted two-state MMPP (hatched) and of the IID two-component exponential mixture (dashed outline), all at $k = 4$. For both FAST sources the observed $\Cmu$ lies far above both synthetic baselines ($3.0\sigma$ and $3.2\sigma$ above the MMPP mean): the inter-session structure exceeds what a two-state rate-switching process can reproduce. \frbc{}'s observed $\Cmu = 0$ is consistent with its baseline ($0.4\sigma$), and its MMPP fit is degenerate (one rate component at $3\%$ stationary occupancy). The $\Delta\mathrm{AIC} = \mathrm{AIC}_\mathrm{MMPP} - \mathrm{AIC}_\mathrm{mix}$ annotation tests whether the MMPP's Markov transitions improve on the IID mixture (negative values favour the MMPP); this model-selection comparison is detailed in Appendix~\ref{app:mmpp_aic}. This plot mirrors Figure~\ref{fig:surrogates}, but the null is different: there it is the memoryless permutation surrogate, here it is a structured rate-switching model fitted to the data, so the observed $\Cmu$ clearing it is the stronger statement that the memory exceeds rate-switching, though part of the excess may reflect the gap-free null's mismatch to the gap-structured data (Section~\ref{sec:literature}).}
    \label{fig:mmpp}
\end{figure*}

If the MMPP captured all the temporal structure, draws from the fitted model would reproduce the observed $\Cmu$. In fact, the \emachine{} captures more structure than the MMPP (Figure~\ref{fig:mmpp}). We summarise each separation as a $\sigma$-equivalent: the number of synthetic standard deviations between the observed $\Cmu$ and the MMPP baseline mean, $z = (\Cmu^\mathrm{obs} - \langle \Cmu^\mathrm{MMPP} \rangle)/\sigma^\mathrm{MMPP}$ (the same standardised effect size used for the surrogate tests in Section~\ref{sec:surrogates}). Because the MMPP $\Cmu$ distributions are non-Gaussian and bounded below at zero, the $\sigma$-equivalent is a descriptive effect size; we therefore also report the empirical tail probability (the fraction of MMPP draws reaching or exceeding the observed $\Cmu$) as the calibrated quantity. For \frba, $\Cmu = 1.116\bits$ is $3.0\sigma$ above the MMPP baseline ($\Cmu^\mathrm{MMPP} = 0.153 \pm 0.323\bits$; only $14$ of $1000$ MMPP draws reach the observed value, an empirical tail probability of $0.014$); for \frbb, $\Cmu = 0.986\bits$ is $3.2\sigma$ above ($\Cmu^\mathrm{MMPP} = 0.111 \pm 0.276\bits$; $6/1000$ draws, $0.006$). For \frbc{} the observed $\Cmu = 0.000\bits$ is consistent with its baseline ($\Cmu^\mathrm{MMPP} = 0.094 \pm 0.253\bits$, $0.4\sigma$), as expected for a source with no detected memory, and its MMPP fit is in any case degenerate (one rate component at 3\% stationary occupancy).

The IID two-component exponential mixture, retained for comparison with the parametric literature, yields larger nominal $\sigma$-separations for the two FAST sources ($3.3\sigma$ for both) because it cannot reproduce any sequential structure by construction; we report it for literature continuity rather than as a substantive null for the sequential claim. The central finding is therefore that the \emachine{} detects temporal structure in the two FAST sources that goes beyond what a two-state rate-switching process can reproduce: the sequential ordering of activity-rate regimes across sessions, not merely the presence of two rate components and their stationary transition structure, carries predictive information. Self-exciting (Hawkes) point processes \citep{Hawkes1971} offer an alternative and widely used description of clustered event sequences, in which each burst transiently raises the rate of subsequent ones; a Hawkes process and a higher-order ($\geq 3$-state) MMPP are the natural next structured nulls against which to test the inter-session $\Cmu$ excess, and we leave them to future work. Detailed reconciliation with the broader FRB temporal-analysis literature (scalar complexity measures, \citealt{SangLin2024, Zhang2024timeenergy}; parametric waiting-time fits, \citealt{Cruces2021}; periodicity searches, \citealt{CHIME2020periodic, Rajwade2020}) is given in Appendices~\ref{app:literature} and \ref{app:scalar}.

\subsection{Implications for observing strategy}
\label{sec:bias_results}

The most prominent feature of our results is the contrast between the FAST sources ($\Cmu \approx 1\bits$, highly significant) and \frbc{} ($\Cmu = 0$, MI consistent with IID at all $k$). Does that contrast reflect the sources' intrinsic physics, or is it an artefact of CHIME's transit-mode observing cadence? To distinguish these, we simulate CHIME-like observations of the two FAST sources: for each FAST session we retain only bursts within the first $\Delta t$ minutes (mimicking a single CHIME transit), recompute the intra-window waiting times, and reconstruct an \emachine{} from the windowed sequence. Table~\ref{tab:windowing} reports the results across window durations $\Delta t = 5, 10, 15, 30, 60$\,min. The simulation reproduces CHIME's window duration but not its higher fluence threshold or per-transit completeness; since those selection effects would further reduce the number of usable waiting times per window, the simulation if anything understates the suppression.

At $\Delta t = 15$\,min, the typical duration of a CHIME transit, $\Cmu$ drops to zero at every $k$ for \frba{} and at every $k$ except $k = 3$ ($\Cmu = 0.810\bits$) for \frbb. At $\Delta t = 60$\,min, \frba's signal is partially recovered ($\Cmu = 0.639\bits$ at $k = 5$; zero at all other $k$). \frbb's $\Delta t = 60$\,min entry is uninformative as a windowing test: its sessions are each only $\sim 56$\,min long, so the 60-min window contains the full session and reproduces the original $\Cmu$ by construction. The scattered non-zero entries at $\Delta t = 10$--$15$\,min ($\Cmu = 0.81$--$0.92\bits$ at isolated $k$ values, vanishing at neighbouring $k$ and $\Delta t$) are consistent with the noise floor of CSSR's $\chi^2$ splitting at small $N$ rather than with recovered signal; no surrogate calibration is attached to any windowed value, so the table should be read for its overall pattern rather than its individual entries.

\frba's partial recovery at $\Delta t = 60$ reflects its session-duration distribution: 39 sessions with median 52\,min but a long tail to 264\,min, with 85\% shorter than 60\,min. The remaining long sessions contribute disproportionately to the waiting-time count and are truncated at $\Delta t = 60$. At 15-minute windows, both sources retain only $\sim 25\%$ of their bursts per session, leaving too few contiguous waiting times for CSSR to detect causal structure. The collapse is therefore consistent with, not in tension with, the inter-session localisation of Section~\ref{sec:one_bit}: windowing leaves the number and ordering of sessions intact but starves the reconstruction of the per-session statistics needed to resolve the regime contrasts, so the signal is lost to statistical power rather than to any removal of inter-session structure.

Reliable detection of temporal memory in the FAST repeaters analysed here therefore appears to require continuous tracking on roughly hour scales. Because the windowed values are single point estimates from an estimator with known bimodal finite-$N$ behaviour, the ${\sim}60$-minute figure is an indicative scale rather than a calibrated threshold, but the overall pattern is clear. Transit-mode surveys, which yield $\sim$15-minute windows per source-day, cannot reliably measure $\Cmu$ even with multi-transit accumulation: $\Cmu$ depends on long contiguous symbol sequences, and concatenating many short transits (as our windowing simulation effectively does across the 39 \frba{} sessions) does not recover the long-range structure. The $\Cmu = 0$ result for \frbc{} is consequently ambiguous: there is an observational degeneracy between genuine memorylessness and insufficient continuous observing duration. The ${\sim}60$-minute scale itself is much longer than the median waiting time (5--40\,s) and comparable to the session duration, consistent with the conclusion of Section~\ref{sec:one_bit} that the predictive structure is carried by the sequence of activity-rate regimes that successive sessions sample. Future studies of FRB temporal complexity should therefore prioritise long-duration tracking observations with instruments capable of hour-scale continuous monitoring.

The status of \frbc{} therefore remains an open question with two possible outcomes. If long-duration tracking observations reveal $\Cmu \approx 1\bits$, the temporal memory would be shared across all three sources analysed here, consistent with a common origin. If $\Cmu$ remains near zero, a genuine dichotomy exists, which could reflect different magnetospheric configurations, magnetic field strengths, progenitor ages, or the same mechanism operating in different regimes. The test does not necessarily require new observations: FAST monitoring of \frbc{} has already yielded more than a thousand bursts over ${\sim}9$\,hr of tracking \citep{Zhang2023fast}, and applying the present analysis to that dataset is the natural next step.

\begin{table*}
    \centering
    \begin{tabular}{@{}lcccccccc@{}}
        \toprule
        & \multicolumn{4}{c}{\frba} & \multicolumn{4}{c}{\frbb} \\
        \cmidrule(lr){2-5} \cmidrule(lr){6-9}
        $\Delta t$ (min) & $k=2$ & $k=3$ & $k=4$ & $k=5$ & $k=2$ & $k=3$ & $k=4$ & $k=5$ \\
        \midrule
        Original & 0.918 & 0.000 & 1.116 & 0.831 & 0.000 & 0.000 & 0.986 & 0.900 \\
        5  & 0.000 & 0.000 & 0.000 & 0.000 & 0.000 & 0.000 & 0.000 & 0.000 \\
        10 & 0.000 & 0.818 & 0.000 & 0.000 & 0.922 & 0.000 & 0.000 & 0.000 \\
        15 & 0.000 & 0.000 & 0.000 & 0.000 & 0.000 & 0.810 & 0.000 & 0.000 \\
        30 & 0.000 & 0.000 & 0.000 & 0.000 & 0.000 & 0.000 & 0.000 & 0.000 \\
        60 & 0.000 & 0.000 & 0.000 & 0.639 & 0.000 & 0.000 & \textbf{0.986} & \textbf{0.900} \\
        \bottomrule
    \end{tabular}
    \caption{Effect of CHIME-like transit-mode windowing on $\Cmu$ (bits). For each FAST source, only bursts within the first $\Delta t$ minutes of each observing session are retained before reconstruction. Bold entries match the original (unwindowed) non-zero value; zero entries that coincide with an original zero are not bolded.}
    \label{tab:windowing}
\end{table*}

\subsection{Limitations}
\label{sec:limitations}

Instrumental systematics from the FAST telescope (time resolution, radio-frequency interference (RFI) flagging, detection threshold effects) are partially controlled by the surrogate testing design: surrogates pass through the same data pipeline and use the same quantile boundaries as the real data, so any systematic that affects only the marginal waiting-time distribution is preserved. An artefact that specifically alters the temporal ordering of bursts without affecting their distribution (for example, time-varying detection sensitivity during high-RFI epochs causing faint bursts to be missed) would not be captured by this test. This caveat carries particular weight because the detected memory is inter-session: any instrumental property that varies from session to session (detection threshold, RFI environment, gain variation with zenith angle, per-session completeness) is in principle degenerate with intrinsic activity-rate regimes. Appendix~\ref{app:fluence_control} tests for such effects directly, repeating the detection on completeness-cut subsamples against a matched-$N$ random-thinning baseline. For \frbb{} the detection survives cuts up to twice the published completeness threshold, so its signal is not carried by bursts near the detection limit. For \frba{} the control is power-limited rather than informative: removing even the 4 per cent of bursts below the \citet{Li2021} completeness threshold collapses the point estimate, but removing the same number of bursts at random does the same in 80 per cent of draws, so the loss reflects the estimator's finite-$N$ firing behaviour rather than evidence of an instrumental origin. A session-level sensitivity covariate check is likewise null once its order-statistics bias is calibrated away. A further consideration bounds the residual risk for \frba{}: \citet{Li2021} correct each burst with a zenith-angle-dependent gain curve and report an off-pulse noise level stable to ${\sim}6$ per cent across the campaign's sessions, leaving little room for threshold drift between sessions. The inter-session attribution should nonetheless be read with the limited power of the completeness control in mind.

The primary methodological limitation is the quantile discretisation step. Converting continuous waiting times to discrete symbols via quantile binning is a lossy projection: some continuous-valued temporal structure may be lost or distorted by the binning. Higher $k$ retains more information (as confirmed by the MI diagnostic, Figure \ref{fig:mi}, which shows increasing MI with $k$) but requires more data for reliable CSSR reconstruction. Our results should therefore be read as estimates of the complexity of the symbolised process at a fixed resolution, not as bounds on the complexity of the underlying continuous process: statistical complexity is not monotone under coarse-graining, and the $k = 5$ overshoot discussed in Section~\ref{sec:cssr_method} shows that the estimator can also exceed the generator's value.

The CSSR hyperparameters and symbolisation scheme introduce a potential methodological sensitivity. $\Cmu$ is insensitive to the significance level $\alpha$ of the state-splitting test, depends on the maximum history length $L$ (\frba{} requires the full $L = 5$, the largest value we sweep, so stability at longer histories is untested; Appendix~\ref{app:hyperparameters}), and is sensitive to the symbolisation scheme: quantile discretisation (our primary method) yields the strongest signal, equal-width binning yields zero for all sources because of bin imbalance for heavy-tailed distributions, and log-spaced binning recovers \frba{}'s signal but not \frbb{}'s. The surrogate tests, which use the same binning as the real data, provide the most reliable inference; the absolute value of $\Cmu$ should be interpreted in the context of the chosen discretisation. Full hyperparameter and binning sensitivity results are reported in Appendix~\ref{app:hyperparameters}.

The bootstrap confidence intervals are wide, reflecting the bimodal nature of $\Cmu$ estimates from finite data. The surrogate test provides a robust statistical assessment, but the absolute value of $\Cmu$ should be interpreted as approximate rather than precise.

CSSR is prone to introducing spurious causal states \citep{Shalizi2004}: the reconstructed machines contain zero-probability states (omitted from Figure~\ref{fig:epsilon_machines}), and the median reconstructed $\Cmu$ overshoots its known value once the $\chi^2$ split test loses calibration at $k = 5$ (Section~\ref{sec:cssr_method}). The number of reconstructed causal states is therefore unreliable, whereas the complexity measure $\Cmu$ is robust: the minimum-$N$ analysis shows it stays near zero for the memoryless Poisson benchmark even where the state topology is not recovered. We therefore do not interpret the number of reconstructed states as a physically meaningful quantity.

The session-boundary control analyses (Section~\ref{sec:one_bit}) and the boundary-free reconstruction (Appendix~\ref{app:boundary_free}) collectively establish that the multi-session $\Cmu$ reported in Section~\ref{sec:emachine_results} reflects inter-session activity-regime switching rather than within-session causal structure of the per-burst emission mechanism. Applying CSSR to the concatenated multi-session sequence operates outside the formal stationarity guarantee of the \emachine{} theorem \citep{Shalizi2001}, so the reported $\Cmu$ for both FAST sources should be interpreted as quantifying inter-session predictive structure rather than the $\Cmu$ of a stationary process.

The within-session-shuffle surrogate test of Appendix~\ref{app:boundary_free} has asymmetric statistical power between the two FAST sources. For \frbb{} (4 sessions, longest with $N = 541$ waiting times above the minimum-sample requirement), the within-session null is well-powered and fails to reject memorylessness; the data are consistent with no within-session predictive memory. For \frba{} (39 sessions, longest with $N = 121$ waiting times below the minimum-sample requirement), the within-session null is power-limited: the pooled within-session-shuffle test does not reject memorylessness after multiple-comparison correction, but borderline raw $p$-values at the finest alphabet resolution and a non-zero $\Cmu = 0.598\bits$ in the longest single session leave room for a weak within-session signal below the detection threshold. We cannot distinguish a genuine absence of within-session memory in \frba{} from one masked by limited per-session sample sizes. This caveat bears directly on \citet{WangWuDai2023}, who report persistence on minute-to-hour (within-session) timescales in both FAST sources via the Hurst exponent: whether that persistence reflects predictive structure that survives our boundary-free and within-session-shuffle controls, rather than the inter-session regime variation we detect, is left to future work (Appendix~\ref{app:literature}).

The windowing analysis shows that $\Cmu$ is sensitive to the duration of continuous observation: sessions shorter than $\sim$60 minutes may fail to detect genuine temporal memory. This places a practical floor on the observing requirements for this technique and limits the number of sources currently amenable to analysis. The windowing test retains the first $\Delta t$ minutes of each session; repeating it with randomly phased or sliding windows, which would more fully bracket the variable transit phasing of a survey such as CHIME, is left to future work, though we expect the $\sim$60-minute detection scale to be robust to that choice.

Finally, our analysis is limited to three sources, of which only two yield conclusive results. A larger sample, enabled by forthcoming data from long-duration tracking campaigns, will be needed to determine whether the temporal memory we detect is universal among repeaters or specific to certain sources.

\section{Conclusions}
\label{sec:conclusions}

We have presented the first application of \emachine{} reconstruction to astrophysical transient timing. By analysing the waiting-time sequences of three repeating FRB sources through the lens of computational mechanics, we have measured the minimum predictive memory, the statistical complexity $\Cmu$, of each source's temporal process. Two FAST-observed sources, \frba{} and \frbb, carry significant temporal memory of order one bit, well above the permutation-surrogate distribution; the absolute value of $\Cmu$ is approximate, but its excess over surrogates is the robust result. \frbc{} (CHIME) shows no detectable memory.

This memory is not a burst-to-burst effect: for both FAST sources it operates on inter-session rather than within-session timescales, reflecting how each source's activity rate varies from one observing session to the next (Section~\ref{sec:one_bit}, Appendix~\ref{app:boundary_free}). The two sources differ in kind: for \frba{} the ordering of sessions is itself predictive, whereas \frbb{}'s signal reflects the contrast between heterogeneous sessions rather than their order. Neither source shows defensible within-session predictive memory, though suggestive evidence persists for \frba{} in its longest session; confirming it would require longer continuous sessions than are currently available.

A simulated observational-bias test shows that CHIME-like transit-mode windowing of FAST data drives $\Cmu$ to zero, so detection of temporal memory requires long continuous observing sessions. The null result for \frbc{} is therefore ambiguous: it may reflect genuinely memoryless emission or an observational limitation of CHIME's short transit windows. Resolving this requires hour-scale continuous tracking of \frbc{}; our windowing analysis predicts that, if \frbc{} possesses memory comparable to the FAST sources, $\Cmu$ should rise above zero when observed for $\gtrsim$\,60~minutes continuously.

The minimal predictive model of each waiting-time sequence requires at least two distinguishable activity-rate regimes, a constraint on the source's activity envelope rather than on the per-burst emission state itself, and the structure exceeds what a two-state Markov-modulated Poisson process can reproduce (Section~\ref{sec:literature}). Because the surviving signal is entirely inter-session, any instrumental property that varies from session to session is in principle degenerate with intrinsic regime switching (Section~\ref{sec:limitations}). Computational mechanics offers a framework for characterising this structure, and forthcoming burst catalogues from CHIME/FRB Catalog~3 and the Deep Synoptic Array \citep[DSA-2000;][]{Hallinan2019} should enable population-level studies of temporal complexity, provided that observing modes yield sufficiently long continuous segments. The framework can also be applied to other astrophysical transients (magnetar burst storms, stellar flares, accretion variability) where the question of temporal memory is equally open.

\section*{Acknowledgements}
This work made use of data from the Five-hundred-meter Aperture Spherical radio Telescope (FAST) and the Canadian Hydrogen Intensity Mapping Experiment (CHIME). We thank the FAST and CHIME/FRB collaborations for making their burst catalogues publicly available.

\section*{Data Availability}

All burst data analysed in this work are public. The FAST burst lists are archived in the Science Data Bank: that of \frba{} \citep{Li2021} at \url{https://doi.org/10.11922/sciencedb.01092}, and that of \frbb{} \citep{Zhang2022} at \url{https://doi.org/10.57760/sciencedb.06762}. The \frbc{} arrival times are from the second CHIME/FRB catalogue \citep{CHIME2026catalog2}, downloaded from the CHIME/FRB public portal at \url{https://www.chime-frb.ca/catalog}.

\section*{Software Availability}

The \emachine{} reconstruction uses the open-source \texttt{emic} package \citep{emic}, available from PyPI at \url{https://pypi.org/project/emic/}. The analysis pipeline developed for this work, including the boundary-free CSSR variant of Appendix~\ref{app:boundary_free}, the surrogate-testing framework, and the scripts that reproduce all tables and figures, will be shared on reasonable request to the corresponding author.

\bibliographystyle{mnras}
\bibliography{references}

@article{Crutchfield1989,
  author  = {Crutchfield, James P. and Young, Karl},
  title   = {Inferring Statistical Complexity},
  journal = {Physical Review Letters},
  volume  = {63},
  number  = {2},
  pages   = {105--108},
  year    = {1989},
  doi     = {10.1103/PhysRevLett.63.105}
}

@article{Shalizi2001,
  author  = {Shalizi, Cosma Rohilla and Crutchfield, James P.},
  title   = {Computational Mechanics: Pattern and Prediction, Structure and Simplicity},
  journal = {Journal of Statistical Physics},
  volume  = {104},
  number  = {3--4},
  pages   = {817--879},
  year    = {2001},
  doi     = {10.1023/A:1010388907793}
}

@article{Crutchfield2012,
  author  = {Crutchfield, James P.},
  title   = {Between Order and Chaos},
  journal = {Nature Physics},
  volume  = {8},
  pages   = {17--24},
  year    = {2012},
  doi     = {10.1038/nphys2190}
}

@phdthesis{Upper1997,
  author = {Upper, Daniel R.},
  title  = {Theory and Algorithms for Hidden {Markov} Models and Generalized Hidden {Markov} Models},
  school = {University of California, Berkeley},
  year   = {1997},
  note   = {Available at \url{https://csc.ucdavis.edu/~cmg/compmech/pubs/TAHMMGHMM.htm}}
}

@article{Marzen2022,
  author  = {Marzen, Sarah E. and Crutchfield, James P.},
  title   = {Inference, Prediction, and Entropy-Rate Estimation of Continuous-Time, Discrete-Event Processes},
  journal = {Entropy},
  volume  = {24},
  number  = {11},
  pages   = {1675},
  year    = {2022},
  doi     = {10.3390/e24111675}
}

@article{Brodu2022,
  author  = {Brodu, Nicolas and Crutchfield, James P.},
  title   = {Discovering Causal Structure with Reproducing-Kernel {Hilbert} Space {$\epsilon$}-Machines},
  journal = {Chaos: An Interdisciplinary Journal of Nonlinear Science},
  volume  = {32},
  number  = {2},
  pages   = {023103},
  year    = {2022},
  doi     = {10.1063/5.0062829}
}

@inproceedings{Shalizi2004,
  author    = {Shalizi, Cosma Rohilla and Shalizi, Kristina Lisa},
  title     = {Blind Construction of Optimal Nonlinear Recursive Predictors for Discrete Sequences},
  booktitle = {Proceedings of the 20th Conference on Uncertainty in Artificial Intelligence (UAI)},
  pages     = {504--511},
  year      = {2004},
  doi       = {10.48550/arXiv.cs/0406011},
  note      = {arXiv:cs/0406011}
}

@article{Shalizi2002,
  author  = {Shalizi, Cosma Rohilla and Shalizi, Kristina Lisa and Crutchfield, James P.},
  title   = {An Algorithm for Pattern Discovery in Time Series},
  journal = {SFI Working Paper 02-10-060},
  year    = {2002},
  doi     = {10.48550/arXiv.cs/0210025},
  note    = {arXiv:cs/0210025}
}

@article{Bartlett2022,
  author  = {Bartlett, Stuart and Li, Jiazheng and Gu, Lixiang and Sinapayen, Lana and Fan, Siteng and Yung, Yuk L.},
  title   = {Assessing Planetary Complexity and Potential Agnostic Biosignatures Using Epsilon Machines},
  journal = {Nature Astronomy},
  volume  = {6},
  pages   = {387--392},
  year    = {2022},
  doi     = {10.1038/s41550-021-01559-x}
}

@article{Suvorova2016,
  author  = {Suvorova, S. and Sun, L. and Melatos, A. and Moran, W. and Evans, R. J.},
  title   = {Hidden {Markov} Model Tracking of Continuous Gravitational Waves from a Neutron Star with Wandering Spin},
  journal = {Physical Review D},
  volume  = {93},
  number  = {12},
  pages   = {123009},
  year    = {2016},
  doi     = {10.1103/PhysRevD.93.123009}
}

@article{Melatos2020,
  author  = {Melatos, A. and Dunn, L. M. and Suvorova, S. and Moran, W. and Evans, R. J.},
  title   = {Pulsar Glitch Detection with a Hidden {Markov} Model},
  journal = {The Astrophysical Journal},
  volume  = {896},
  number  = {1},
  pages   = {78},
  year    = {2020},
  doi     = {10.3847/1538-4357/ab9178}
}

@article{Abbott2022ScoX1,
  author  = {Abbott, R. and others},
  title   = {Search for Gravitational Waves from {Scorpius~X-1} with a Hidden {Markov} Model in {O3} {LIGO} Data},
  journal = {Physical Review D},
  volume  = {106},
  number  = {6},
  pages   = {062002},
  year    = {2022},
  doi     = {10.1103/PhysRevD.106.062002}
}

@article{Zimmerman2024,
  author  = {Zimmerman, Robert and van Dyk, David A. and Kashyap, Vinay L. and Siemiginowska, Aneta},
  title   = {Separating States in Astronomical Sources Using Hidden {Markov} Models: With a Case Study of Flaring and Quiescence on {EV~Lac}},
  journal = {Monthly Notices of the Royal Astronomical Society},
  volume  = {534},
  number  = {3},
  pages   = {2142--2167},
  year    = {2024},
  doi     = {10.1093/mnras/stae2082}
}

@article{Kimpson2024Kalman,
  author  = {Kimpson, Tom and Melatos, Andrew and O'Leary, Joseph and Carlin, Julian B. and Evans, Robin J. and Moran, William and Cheunchitra, Tong and Dong, Wenhao and Dunn, Liam and Greentree, Julian and O'Neill, Nicholas J. and Suvorova, Sofia and Thong, Kok Hong and Vargas, Andr{\'e}s F.},
  title   = {Kalman Tracking and Parameter Estimation of Continuous Gravitational Waves with a Pulsar Timing Array},
  journal = {Monthly Notices of the Royal Astronomical Society},
  volume  = {534},
  number  = {3},
  pages   = {1844--1867},
  year    = {2024},
  doi     = {10.1093/mnras/stae2197}
}

@article{Kimpson2024PulsarTerms,
  author  = {Kimpson, Tom and Melatos, Andrew and O'Leary, Joseph and Carlin, Julian B. and Evans, Robin J. and Moran, William and Cheunchitra, Tong and Dong, Wenhao and Dunn, Liam and Greentree, Julian and O'Neill, Nicholas J. and Suvorova, Sofia and Thong, Kok Hong and Vargas, Andr{\'e}s F.},
  title   = {State-space Analysis of a Continuous Gravitational Wave Source with a Pulsar Timing Array: Inclusion of the Pulsar Terms},
  journal = {Monthly Notices of the Royal Astronomical Society},
  volume  = {535},
  number  = {1},
  pages   = {132--154},
  year    = {2024},
  doi     = {10.1093/mnras/stae2360}
}

@article{Kimpson2025Background,
  author  = {Kimpson, Tom and Melatos, Andrew and O'Leary, Joseph and Carlin, Julian B. and Evans, Robin J. and Moran, William and Cheunchitra, Tong and Dong, Wenhao and Dunn, Liam and Greentree, Julian and O'Neill, Nicholas J. and Suvorova, Sofia and Thong, Kok Hong and Vargas, Andr{\'e}s F.},
  title   = {State-space Algorithm for Detecting the Nanohertz Gravitational Wave Background},
  journal = {Monthly Notices of the Royal Astronomical Society},
  volume  = {537},
  number  = {2},
  pages   = {1282--1304},
  year    = {2025},
  doi     = {10.1093/mnras/staf068}
}

@article{Kimpson2024Mains,
  author  = {Kimpson, Tom and Suvorova, Sofia and Middleton, Hannah and Liu, Changrong and Melatos, Andrew and Evans, Robin J. and Moran, William},
  title   = {Adaptive Cancellation of Mains Power Interference in Continuous Gravitational Wave Searches with a Hidden {Markov} Model},
  journal = {Physical Review D},
  volume  = {110},
  number  = {12},
  pages   = {122004},
  year    = {2024},
  doi     = {10.1103/PhysRevD.110.122004}
}

@article{OLeary2024UKF,
  author  = {O'Leary, Joseph and Melatos, Andrew and Kimpson, Tom and O'Neill, Nicholas J. and Meyers, Patrick M. and Christodoulou, Dimitris M. and Bhattacharya, Sayantan and Laycock, Silas G. T.},
  title   = {Measuring the Magnetic Dipole Moment and Magnetospheric Fluctuations of Accretion-powered Pulsars in the Small Magellanic Cloud with an Unscented {Kalman} Filter},
  journal = {The Astrophysical Journal},
  volume  = {971},
  number  = {2},
  pages   = {126},
  year    = {2024},
  doi     = {10.3847/1538-4357/ad53c2}
}

@article{OLeary2025RT,
  author  = {O'Leary, Joseph and Melatos, Andrew and Kimpson, Tom and Christodoulou, Dimitris M. and O'Neill, Nicholas J. and Meyers, Patrick M. and Bhattacharya, Sayantan and Laycock, Silas G. T.},
  title   = {Observing Rayleigh--Taylor Stable and Unstable Accretion through a {Kalman} Filter Analysis of X-ray Pulsars in the Small Magellanic Cloud},
  journal = {The Astrophysical Journal},
  volume  = {981},
  number  = {2},
  pages   = {150},
  year    = {2025},
  doi     = {10.3847/1538-4357/adaf26}
}

@article{OLeary2025Torque,
  author  = {O'Leary, Joseph and Melatos, Andrew and Kimpson, Tom and Christodoulou, Dimitris M. and O'Neill, Nicholas J. and Meyers, Patrick M. and Bhattacharya, Sayantan and Laycock, Silas G. T.},
  title   = {Is There a Retrograde Accretion Disk around {4U~1626-67}? Tracking Torque Reversals with a State-space Model},
  journal = {The Astrophysical Journal},
  volume  = {997},
  number  = {1},
  pages   = {130},
  year    = {2026},
  doi     = {10.3847/1538-4357/ae2610}
}

@article{OLeary2026HMM,
  author  = {O'Leary, Joseph and Dunn, Liam and Melatos, Andrew},
  title   = {Discovering Pulsars in Compact Binaries with a Hidden {Markov} Model},
  journal = {The Astrophysical Journal},
  volume  = {998},
  number  = {1},
  pages   = {183},
  year    = {2026},
  doi     = {10.3847/1538-4357/ae3288}
}

@article{TotaniTsuzuki2023,
  author  = {Totani, Tomonori and Tsuzuki, Yuya},
  title   = {Fast Radio Bursts Trigger Aftershocks Resembling Earthquakes, but not Solar Flares},
  journal = {Monthly Notices of the Royal Astronomical Society},
  volume  = {526},
  number  = {2},
  pages   = {2795--2811},
  year    = {2023},
  doi     = {10.1093/mnras/stad2532}
}

@article{Zhang2024timeenergy,
  author  = {Zhang, Yong-Kun and Li, Di and Feng, Yi and Wang, Pei and Niu, Chen-Hui and Dai, Shi and Yao, Ju-Mei and Tsai, Chao-Wei},
  title   = {The Arrival Time and Energy of {FRBs} Traverse the Time--Energy Bivariate Space Like a {Brownian} Motion},
  journal = {Science Bulletin},
  volume  = {69},
  number  = {8},
  pages   = {1020--1026},
  year    = {2024},
  doi     = {10.1016/j.scib.2024.02.010}
}

@article{SangLin2024,
  author  = {Sang, Yu and Lin, Hai-Nan},
  title   = {Quantifying the Randomness and Scale Invariance of the Repeating Fast Radio Bursts},
  journal = {Monthly Notices of the Royal Astronomical Society},
  volume  = {533},
  number  = {1},
  pages   = {872--879},
  year    = {2024},
  doi     = {10.1093/mnras/stae1873}
}

@article{Cruces2021,
  author  = {Cruces, M. and Spitler, L. G. and Scholz, P. and others},
  title   = {Repeating Behaviour of {FRB} 121102: Periodicity, Waiting Times, and Energy Distribution},
  journal = {Monthly Notices of the Royal Astronomical Society},
  volume  = {500},
  number  = {1},
  pages   = {448--463},
  year    = {2021},
  doi     = {10.1093/mnras/staa3223}
}

@article{Oppermann2018,
  author  = {Oppermann, Niels and Yu, Hao-Ran and Pen, Ue-Li},
  title   = {On the Non-{Poissonian} Repetition Pattern of {FRB} 121102},
  journal = {Monthly Notices of the Royal Astronomical Society},
  volume  = {475},
  number  = {4},
  pages   = {5109--5115},
  year    = {2018},
  doi     = {10.1093/mnras/sty004}
}

@article{Rajwade2020,
  author  = {Rajwade, K. M. and Mickaliger, M. B. and Stappers, B. W. and others},
  title   = {Possible Periodic Activity in the Repeating {FRB} 121102},
  journal = {Monthly Notices of the Royal Astronomical Society},
  volume  = {495},
  number  = {4},
  pages   = {3551--3558},
  year    = {2020},
  doi     = {10.1093/mnras/staa1237}
}

@article{Gourdji2019,
  author  = {Gourdji, K. and Michilli, D. and Spitler, L. G. and others},
  title   = {A Sample of Low-Energy Bursts from {FRB}~121102},
  journal = {The Astrophysical Journal Letters},
  volume  = {877},
  number  = {2},
  pages   = {L19},
  year    = {2019},
  doi     = {10.3847/2041-8213/ab1f8a}
}

@article{Aggarwal2021,
  author  = {Aggarwal, K. and Agarwal, D. and Lewis, E. F. and others},
  title   = {Comprehensive Analysis of a Dense Sample of {FRB} 121102 Bursts},
  journal = {The Astrophysical Journal},
  volume  = {922},
  number  = {2},
  pages   = {115},
  year    = {2021},
  doi     = {10.3847/1538-4357/ac2577}
}

@article{Jahns2023,
  author  = {Jahns, J. N. and Spitler, L. G. and Nimmo, K. and others},
  title   = {The {FRB}~20121102{A} November Rain in 2018 Observed with the {Arecibo} Telescope},
  journal = {Monthly Notices of the Royal Astronomical Society},
  volume  = {519},
  number  = {1},
  pages   = {666--687},
  year    = {2023},
  doi     = {10.1093/mnras/stac3446}
}

@article{WangWuDai2023,
  author  = {Wang, F. Y. and Wu, Q. and Dai, Z. G.},
  title   = {Repeating Fast Radio Bursts Reveal Memory from Minutes to an Hour},
  journal = {The Astrophysical Journal Letters},
  volume  = {949},
  number  = {2},
  pages   = {L33},
  year    = {2023},
  doi     = {10.3847/2041-8213/acd5d2}
}

@article{Wang2024memory,
  author  = {Wang, Ping and Song, Li-Ming and Xiong, Shao-Lin and others},
  title   = {Memory in the Burst Occurrence of Repeating {FRBs}},
  journal = {The Astrophysical Journal},
  volume  = {975},
  number  = {2},
  pages   = {188},
  year    = {2024},
  doi     = {10.3847/1538-4357/ad7de5}
}

@article{Du2024scaling,
  author  = {Du, Yan-Qi and Wang, Ping and Song, Li-Ming and Xiong, Shao-Lin},
  title   = {Scaling and Universality in the Temporal Occurrence of Repeating {FRBs}},
  journal = {Monthly Notices of the Royal Astronomical Society: Letters},
  volume  = {531},
  number  = {1},
  pages   = {L57--L62},
  year    = {2024},
  doi     = {10.1093/mnrasl/slae031}
}

@article{Li2021,
  author  = {Li, D. and Wang, P. and Zhu, W. W. and others},
  title   = {A Bimodal Burst Energy Distribution of a Repeating Fast Radio Burst Source},
  journal = {Nature},
  volume  = {598},
  pages   = {267--271},
  year    = {2021},
  doi     = {10.1038/s41586-021-03878-5}
}

@article{Zhang2022,
  author  = {Zhang, Y.-K. and Wang, P. and Feng, Y. and Zhang, B. and Li, D. and Tsai, C.-W. and Niu, C.-H. and Luo, R. and Yao, J.-M. and Zhu, W.-W. and Han, J.-L. and Lee, K.-J. and Zhou, D.-J. and Niu, J.-R. and Jiang, J.-C. and Wang, W.-Y. and Zhang, C.-F. and Xu, H. and Wang, B.-J. and Xu, J.-W.},
  title   = {{FAST} Observations of an Extremely Active Episode of {FRB} 20201124A. {II}. {Energy} Distribution},
  journal = {Research in Astronomy and Astrophysics},
  volume  = {22},
  number  = {12},
  pages   = {124002},
  year    = {2022},
  doi     = {10.1088/1674-4527/ac98f7},
  eprint  = {2210.03645}
}

@article{Xu2022,
  author  = {Xu, Heng and Niu, J. R. and Chen, Peng and others},
  title   = {A Fast Radio Burst Source at a Complex Magnetized Site in a Barred Galaxy},
  journal = {Nature},
  volume  = {609},
  pages   = {685--688},
  year    = {2022},
  doi     = {10.1038/s41586-022-05071-8}
}

@article{CHIME2020periodic,
  author  = {{CHIME/FRB Collaboration}},
  title   = {Periodic Activity from a Fast Radio Burst Source},
  journal = {Nature},
  volume  = {582},
  pages   = {351--355},
  year    = {2020},
  doi     = {10.1038/s41586-020-2398-2}
}

@article{CHIME2026catalog2,
  author  = {{CHIME/FRB Collaboration}},
  title   = {The Second {CHIME/FRB} Catalog of Fast Radio Bursts},
  journal = {The Astrophysical Journal Supplement Series},
  year    = {2026},
  volume  = {283},
  number  = {1},
  pages   = {34},
  doi     = {10.3847/1538-4365/ae3828},
  note    = {arXiv:2601.09399}
}

@article{Zhang2023review,
  author  = {Zhang, Bing},
  title   = {The Physics of Fast Radio Bursts},
  journal = {Reviews of Modern Physics},
  volume  = {95},
  pages   = {035005},
  year    = {2023},
  doi     = {10.1103/RevModPhys.95.035005}
}

@article{Petroff2022,
  author  = {Petroff, E. and Hessels, J. W. T. and Lorimer, D. R.},
  title   = {Fast Radio Bursts at the Dawn of the 2020s},
  journal = {The Astronomy and Astrophysics Review},
  volume  = {30},
  pages   = {2},
  year    = {2022},
  doi     = {10.1007/s00159-022-00139-w}
}

@article{Wadiasingh2019,
  author  = {Wadiasingh, Z. and Timokhin, A.},
  title   = {Repeating Fast Radio Bursts from Magnetars with Low Magnetospheric Twist},
  journal = {The Astrophysical Journal},
  volume  = {879},
  pages   = {4},
  year    = {2019},
  doi     = {10.3847/1538-4357/ab2240}
}

@article{Lu2020,
  author  = {Lu, Wenbin and Kumar, Pawan and Zhang, Bing},
  title   = {A Unified Picture of {Galactic} and Cosmological Fast Radio Bursts},
  journal = {Monthly Notices of the Royal Astronomical Society},
  volume  = {498},
  pages   = {1397--1405},
  year    = {2020},
  doi     = {10.1093/mnras/staa2450}
}

@article{Beloborodov2020,
  author  = {Beloborodov, A. M.},
  title   = {Blast Waves from Magnetar Flares and Fast Radio Bursts},
  journal = {The Astrophysical Journal},
  volume  = {896},
  pages   = {142},
  year    = {2020},
  doi     = {10.3847/1538-4357/ab83eb}
}

@article{Lorimer2007,
  author  = {Lorimer, D. R. and Bailes, M. and McLaughlin, M. A. and Narkevic, D. J. and Crawford, F.},
  title   = {A Bright Millisecond Radio Burst of Extragalactic Origin},
  journal = {Science},
  volume  = {318},
  pages   = {777--780},
  year    = {2007},
  doi     = {10.1126/science.1147532}
}

@article{Spitler2016,
  author  = {Spitler, L. G. and Scholz, P. and Hessels, J. W. T. and others},
  title   = {A Repeating Fast Radio Burst},
  journal = {Nature},
  volume  = {531},
  pages   = {202--205},
  year    = {2016},
  doi     = {10.1038/nature17168}
}

@article{Platts2019,
  author  = {Platts, E. and Weltman, A. and Walters, A. and others},
  title   = {A Living Theory Catalogue for Fast Radio Bursts},
  journal = {Physics Reports},
  volume  = {821},
  pages   = {1--27},
  year    = {2019},
  doi     = {10.1016/j.physrep.2019.06.003}
}

@article{CHIME2020nature,
  author  = {{CHIME/FRB Collaboration}},
  title   = {A Bright Millisecond-Duration Radio Burst from a {Galactic} Magnetar},
  journal = {Nature},
  volume  = {587},
  pages   = {54--58},
  year    = {2020},
  doi     = {10.1038/s41586-020-2863-y}
}

@article{Bochenek2020,
  author  = {Bochenek, C. D. and Ravi, V. and Belov, K. V. and others},
  title   = {A Fast Radio Burst Associated with a {Galactic} Magnetar},
  journal = {Nature},
  volume  = {587},
  pages   = {59--62},
  year    = {2020},
  doi     = {10.1038/s41586-020-2872-x}
}

@book{CoverThomas2006,
  author    = {Cover, Thomas M. and Thomas, Joy A.},
  title     = {Elements of Information Theory},
  edition   = {2nd},
  publisher = {Wiley-Interscience},
  year      = {2006}
}

@article{Pincus1991,
  author  = {Pincus, Steven M.},
  title   = {Approximate Entropy as a Measure of System Complexity},
  journal = {Proceedings of the National Academy of Sciences},
  volume  = {88},
  number  = {6},
  pages   = {2297--2301},
  year    = {1991},
  doi     = {10.1073/pnas.88.6.2297}
}

@article{Hawkes1971,
  author  = {Hawkes, Alan G.},
  title   = {Spectra of Some Self-Exciting and Mutually Exciting Point Processes},
  journal = {Biometrika},
  volume  = {58},
  number  = {1},
  pages   = {83--90},
  year    = {1971},
  doi     = {10.1093/biomet/58.1.83}
}

@article{Theiler1992,
  author  = {Theiler, James and Eubank, Stephen and Longtin, Andr{\'e} and Galdrikian, Bryan and Farmer, J. Doyne},
  title   = {Testing for Nonlinearity in Time Series: The Method of Surrogate Data},
  journal = {Physica D: Nonlinear Phenomena},
  volume  = {58},
  number  = {1--4},
  pages   = {77--94},
  year    = {1992},
  doi     = {10.1016/0167-2789(92)90102-S}
}

@article{Shannon1948,
  author  = {Shannon, Claude E.},
  title   = {A Mathematical Theory of Communication},
  journal = {The Bell System Technical Journal},
  volume  = {27},
  number  = {3},
  pages   = {379--423},
  year    = {1948},
  doi     = {10.1002/j.1538-7305.1948.tb01338.x}
}

@article{Kantelhardt2002,
  author  = {Kantelhardt, J. W. and Zschiegner, S. A. and Koscielny-Bunde, E. and Havlin, S. and Bunde, A. and Stanley, H. E.},
  title   = {Multifractal detrended fluctuation analysis of nonstationary time series},
  journal = {Physica A: Statistical Mechanics and its Applications},
  volume  = {316},
  pages   = {87--114},
  year    = {2002},
  doi     = {10.1016/S0378-4371(02)01383-3}
}

@misc{emic,
  author = {Azariah, John},
  title  = {emic: A {Python} framework for constructing and analyzing epsilon-machines based on computational mechanics},
  year   = {2026},
  url    = {https://pypi.org/project/emic/},
  note   = {Python package}
}

@article{Efron1979,
  author  = {Efron, Bradley},
  title   = {Bootstrap Methods: Another Look at the Jackknife},
  journal = {The Annals of Statistics},
  volume  = {7},
  number  = {1},
  pages   = {1--26},
  year    = {1979},
  doi     = {10.1214/aos/1176344552}
}

@article{Kunsch1989,
  author  = {K{\"u}nsch, Hans R.},
  title   = {The Jackknife and the Bootstrap for General Stationary Observations},
  journal = {The Annals of Statistics},
  volume  = {17},
  number  = {3},
  pages   = {1217--1241},
  year    = {1989},
  doi     = {10.1214/aos/1176347265}
}

@article{HallHorowitzJing1995,
  author  = {Hall, Peter and Horowitz, Joel L. and Jing, Bing-Yi},
  title   = {On Blocking Rules for the Bootstrap with Dependent Data},
  journal = {Biometrika},
  volume  = {82},
  number  = {3},
  pages   = {561--574},
  year    = {1995},
  doi     = {10.1093/biomet/82.3.561}
}

@article{Hewitt2022,
  author  = {Hewitt, D. M. and Snelders, M. P. and Hessels, J. W. T. and others},
  title   = {Arecibo Observations of a Burst Storm from {FRB} 20121102A in 2016},
  journal = {Monthly Notices of the Royal Astronomical Society},
  volume  = {515},
  pages   = {3577--3596},
  year    = {2022},
  doi     = {10.1093/mnras/stac1960}
}

@article{SchreiberSchmitz1996,
  author  = {Schreiber, Thomas and Schmitz, Andreas},
  title   = {Improved Surrogate Data for Nonlinearity Tests},
  journal = {Physical Review Letters},
  volume  = {77},
  number  = {4},
  pages   = {635--638},
  year    = {1996},
  doi     = {10.1103/PhysRevLett.77.635}
}

@article{BenjaminiHochberg1995,
  author  = {Benjamini, Yoav and Hochberg, Yosef},
  title   = {Controlling the False Discovery Rate: A Practical and Powerful Approach to Multiple Testing},
  journal = {Journal of the Royal Statistical Society: Series B (Methodological)},
  volume  = {57},
  number  = {1},
  pages   = {289--300},
  year    = {1995},
  doi     = {10.1111/j.2517-6161.1995.tb02031.x}
}

@article{PhipsonSmyth2010,
  author  = {Phipson, Belinda and Smyth, Gordon K.},
  title   = {Permutation P-values Should Never Be Zero: Calculating Exact P-values When Permutations Are Randomly Drawn},
  journal = {Statistical Applications in Genetics and Molecular Biology},
  volume  = {9},
  number  = {1},
  pages   = {Article 39},
  year    = {2010},
  doi     = {10.2202/1544-6115.1585}
}

@article{Hallinan2019,
  author  = {Hallinan, G. and Ravi, V. and Weinreb, S. and others},
  title   = {The {DSA}-2000 --- A Radio Survey Camera},
  journal = {Bulletin of the American Astronomical Society},
  volume  = {51},
  pages   = {255},
  year    = {2019},
  note    = {arXiv:1907.07648}
}

@article{Zhang2023fast,
  author  = {Zhang, Yong-Kun and Li, Di and Zhang, Bing and others},
  title   = {{FAST} Observations of {FRB} 20220912A: Burst Properties and Polarization Characteristics},
  journal = {The Astrophysical Journal},
  volume  = {955},
  number  = {2},
  pages   = {142},
  year    = {2023},
  doi     = {10.3847/1538-4357/aced0b}
}

\appendix

\section{Synthetic Validation}
\label{app:validation}

Before applying the pipeline to real FRB data, we validate it on synthetic processes with known causal structure.

\subsection{Known \emachine{} recovery}
\label{sec:known_machines}

We first verify that the \texttt{emic} CSSR implementation correctly recovers analytically known \emachines. The Even Process (1s occur only in blocks of even length, bounded by 0s) and the Golden Mean Process (no consecutive 1s) are two canonical two-state \emachines{}. Under fair-coin emission from the stochastic state (the standard convention in \citet{Crutchfield2012}, which we adopt), both processes have stationary distribution $\pi = (2/3, 1/3)$, statistical complexity $\Cmu = \log_{2} 3 - 2/3 \approx 0.918\bits$, and entropy rate $\hmu = 2/3 \approx 0.667\bitssym$. They differ only in which symbol the deterministic state must emit: ``1'' for the Even Process (completing an even-length block) and ``0'' for the Golden Mean Process (forbidding a second consecutive 1). The two are thus distinct \emachine{} topologies with identical complexity, so recovering both separately tests the topology and not merely the scalar $\Cmu$. From $10\,000$-symbol sequences, \texttt{emic} recovers both machines with the correct number of states and $\Cmu$ within $0.5\%$ of the analytical value.

\subsection{End-to-end pipeline validation}
\label{sec:pipeline_validation}

We test the full pipeline (waiting-time generation, quantile symbolisation, CSSR reconstruction, and complexity measurement) on two synthetic processes. A homogeneous Poisson process (rate $\lambda = 1$\,s$^{-1}$) represents the memoryless null model; a two-state HMM with distinct burst rates in each state represents a structured process with known causal complexity.

The Poisson process yields $\Cmu \approx 0$, $\hmu = 1.58\bitssym$, and the best-fit waiting-time distribution is exponential, as expected; $\hmu = \log_2 3$, the exact value for an IID sequence quantile-binned into $k = 3$ equally occupied symbols, which provides a closed-form check. The two-state HMM yields $\Cmu = 0.96\bits$, $\hmu = 1.35\bitssym$, and approximate entropy (ApEn; \citealt{Pincus1991}) $= 0.45$ (compared to 1.69 for the Poisson process). The reconstructed $\Cmu = 0.96\bits$ slightly exceeds the generator's two-state hidden-state entropy ($H(\pi) = 0.918\bits$; Section~\ref{sec:minimum_n}) because the \emachine{} of a non-unifilar HMM may resolve additional causal states (Section~\ref{sec:hmm_relation}), so the two quantities need not coincide. The pipeline correctly identifies the Poisson process as memoryless and the HMM as structured.

\subsection{Minimum-$N$ analysis}
\label{sec:minimum_n}

The quantity that matters for the real-data inference is not the bias of the $\Cmu$ point estimate but the behaviour of the test we actually apply to it: at a given sequence length, how often does the permutation-surrogate procedure of Section~\ref{sec:surrogates} flag genuine structure, and how often does it fire on a memoryless process? We characterise both by passing two synthetic processes of known character through the identical symbolise--reconstruct--surrogate pipeline used on the FRB data: a homogeneous Poisson process (memoryless) and the two-state HMM of Section~\ref{sec:pipeline_validation} (structured, with hidden-state entropy $H(\pi) = 0.918\bits$). For each process we generate sequences of length $N = 100, 200, 500, 1000, 2000$ (200 independent realisations per $N$), sweep the alphabet size $k = 2$--$5$ at $L = 5$, and apply the one-sided permutation test (200 surrogates, rejection level $0.05$) to every realisation. The fraction of HMM realisations rejecting the memoryless null is the statistical power; the same fraction for the Poisson realisations is the empirical false-positive rate.

\begin{figure}
    \centering
    \includegraphics[width=\columnwidth]{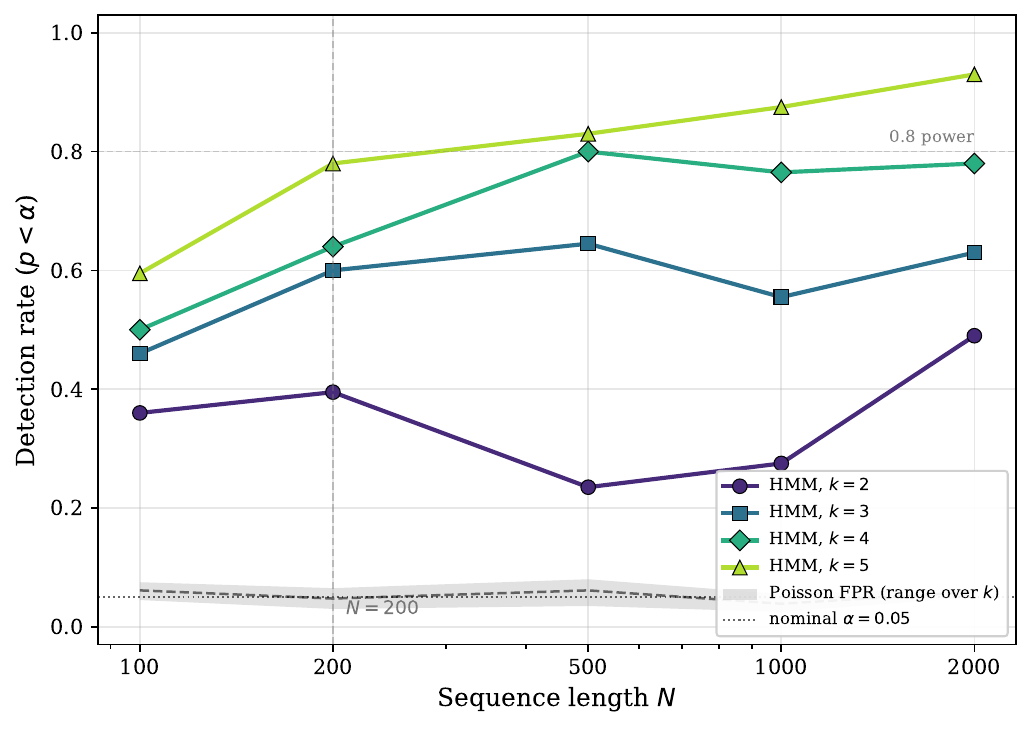}
    \caption{Detection power of the permutation-surrogate test (Section~\ref{sec:surrogates}) as a function of sequence length $N$, from the synthetic validation. Coloured curves give the fraction of 200 structured (two-state HMM) realisations for which the one-sided test rejects the memoryless null at the $0.05$ level (i.e. the statistical power) at alphabet sizes $k = 2$--$5$. The grey band spans the corresponding rejection fraction for memoryless (Poisson) realisations across all $k$, i.e.\ the empirical false-positive rate, which stays close to the nominal $0.05$ level (dotted line) at every $N$, confirming the test is well calibrated. Power is set more by alphabet resolution than by $N$ over this range, reaching ${\sim}0.78$ at the fiducial $k = 4$ and ${\sim}0.88$ at $k = 5$ for $N \gtrsim 500$; the shortfall at low $k$ is a coarse-resolution reconstruction failure (Section~\ref{sec:mi_diagnostic}), not a sample-size effect. The vertical dashed line marks the $N = 200$ minimum-sample requirement adopted for source selection (Section~\ref{sec:data}). Curves reflect 200-realisation sampling noise.}
    \label{fig:synthetic_validation}
\end{figure}

The false-positive rate tracks the nominal level closely (Figure~\ref{fig:synthetic_validation}, grey band): it stays at or below $0.08$ at every $k$ and $N$, with a mean of $0.05$. The permutation test is therefore well calibrated, and the elevated $\Cmu$ we report for the FAST sources is not an artefact of the reconstruction procedure. Detection power, by contrast, is governed more by the alphabet resolution $k$ than by $N$ over this range. At $k = 2$ the test has little power (never exceeding ${\sim}0.5$); power rises with resolution, reaching ${\sim}0.64$--$0.80$ at the fiducial $k = 4$ and ${\sim}0.78$--$0.93$ at $k = 5$ for $N \gtrsim 200$. The shortfall from unity at low $k$ is a coarse-resolution reconstruction failure rather than a sample-size effect: a fraction of structured realisations collapse to a trivial single-state machine ($\Cmu = 0$) because the conservative $\chi^2$ split test cannot separate the two rate regimes once they are merged into a coarse alphabet. This fraction falls from ${\sim}0.5$--$0.75$ at $k = 2$ to ${\sim}0.1$ at $k = 4$--$5$ and is essentially independent of $N$, mirroring the coarse-resolution $\Cmu = 0$ entries seen in the real data and diagnosed by the mutual-information test of Section~\ref{sec:mi_diagnostic}.

Separately, $\Cmu$ is a more robust discriminator than the reconstructed state count. CSSR is prone to introducing spurious states \citep{Shalizi2004}, but $\Cmu$ remains near zero for the Poisson process because such states carry near-identical emission distributions; the complexity measure is stable even where the state topology is not. This is why we report and interpret $\Cmu$ rather than the number of reconstructed states.

Two practical conclusions follow. First, $N \gtrsim 200$ is a pragmatic floor rather than a sharp detection threshold. Power is already moderate at $N = 100$ (e.g.\ ${\sim}0.5$ at $k = 4$) and improves only gradually thereafter. What changes across $N \sim 100$--$200$ is that power becomes more uniform across resolutions: at $N = 100$ only $k = 5$ reaches ${\sim}0.6$, whereas by $N = 200$ all of $k = 3$--$5$ reach it. The $\chi^2$ state-splitting test also stays better calibrated, with expected cell counts ${\sim}N/k^{L+1}$, so halving $N$ pushes even the fiducial $k = 4$ below the marginal regime of Section~\ref{sec:cssr_method}. Because the synthetic HMM is a clean, well-separated best case, real burst sequences are likely to be detected less readily at a given $N$, motivating the more conservative floor. Both FAST sources sit far above it ($N_\mathrm{wt} = 1613$ and $877$), whereas \frbc{} ($N_\mathrm{wt} = 193$) sits just below it, so its null result (Section~\ref{sec:emachine_results}) is consistent with either genuine memorylessness or a missed detection, which reinforces the windowing analysis of Section~\ref{sec:bias_results}. Second, because the point estimate is noisy and false positives are controlled only by the surrogate procedure rather than by thresholding $\Cmu$ directly, all inference on the real data rests on the permutation test and block bootstrap, not on the bare $\Cmu$ value.

\section{Boundary-Free Reconstruction and Within-Session Surrogates}
\label{app:boundary_free}

The canonical analysis concatenates intra-session waiting times into a single 1-D array before passing them to CSSR. With $L = 5$, the resulting suffix tree includes histories whose $(L+1)$-symbol span crosses between successive sessions: there are at most $L$ such spurious histories per session boundary. For \frba{} (38 boundaries across 1613 intra-session waiting times) this affects $\leq 190$ histories (${\sim}12\%$ of the data); for \frbb{} (3 boundaries across 877 waiting times) only $\leq 15$ histories (${\sim}1.7\%$). To establish how much of the canonical $\Cmu$ depends on these cross-session histories, we perform a boundary-free reconstruction that builds the CSSR suffix tree per session, and test both the boundary-free and canonical results against a within-session-shuffle null surrogate that preserves per-session waiting-time multisets and destroys only within-session temporal ordering.

\subsection{Boundary-free CSSR reconstruction}
\label{sec:boundary_free_method}

The canonical implementation calls \texttt{emic}'s \texttt{SuffixTree.build\_from\_sequence}, which slides a window of length $L$ across the concatenated symbol array and registers each (history, next-symbol) pair into the tree. The boundary-free variant replaces this single call with a loop over per-session symbol sequences, calling the lower-level \texttt{SuffixTree.add\_observation} method only for positions where both the history of length $\ell \leq L$ and its next symbol lie within the same session. All subsequent CSSR phases (sufficiency, determinism, machine construction) are run unchanged on the resulting boundary-free suffix tree. Symbolisation uses the same global quantile boundaries as the canonical analysis (Section~\ref{sec:symbolisation}), so the marginal symbol distribution is identical at both levels. A session that yields only a single waiting time forms no (history, next-symbol) pair and so contributes nothing to the boundary-free tree; \frba{} has two such sessions (two waiting times in total), \frbb{} none.

Table~\ref{tab:boundary_free_ksweep} reports the boundary-free $\Cmu$ alongside the canonical value across the same $k$-sweep used in Table~\ref{tab:surrogates}. For \frbb{} the boundary-free reconstruction is essentially indistinguishable from the canonical at every $k$, as expected given the small number of session boundaries. For \frba{}, the boundary-free reconstruction collapses to $\Cmu = 0$ at $k = 2, 3, 4$ (including at the canonical fiducial $k = 4$, where the concatenated $\Cmu = 1.116\bits$) and yields $\Cmu = 1.070\bits$ at $k = 5$. The $k = 5$ result is approximate (the $\chi^2$ null is not formally justified at this resolution; see Section~\ref{sec:cssr_method}) and is interpreted alongside the surrogate-test result below rather than on its own. That the boundary-free value at $k = 5$ exceeds the canonical $0.831\bits$ despite the boundary-free tree containing strictly fewer histories is itself a symptom of this uncalibrated regime: removing sparse cross-session histories changes which marginal splits the $\chi^2$ test makes at cell counts far below calibration, and the resulting state structure (and hence $\Cmu$) can move in either direction.

\begin{table}
    \centering
    \begin{tabular}{@{}lcccc@{}}
        \toprule
        Source & $k=2$ & $k=3$ & $k=4$ & $k=5$ \\
        \midrule
        \multicolumn{5}{l}{\emph{Canonical (concatenated)}} \\
        \frba & 0.918 & 0.000 & 1.116 & 0.831 \\
        \frbb & 0.000 & 0.000 & 0.986 & 0.900 \\
        \multicolumn{5}{l}{\emph{Boundary-free}} \\
        \frba & 0.000 & 0.000 & 0.000 & 1.070 \\
        \frbb & 0.000 & 0.000 & 0.985 & 0.895 \\
        \bottomrule
    \end{tabular}
    \caption{Canonical (concatenated) and boundary-free $\Cmu$ (bits) for the two FAST sources at $L = 5$, $\alpha = 0.001$. The boundary-free reconstruction builds CSSR's suffix tree per session so that no history of length $L$ spans a session boundary.}
    \label{tab:boundary_free_ksweep}
\end{table}

\subsection{Within-session-shuffle surrogates}
\label{sec:within_session_surr}

The permutation surrogates of Section~\ref{sec:emachine_results} shuffle waiting times across the entire concatenated sequence, destroying both within- and inter-session temporal ordering. They therefore test the null hypothesis ``the waiting-time sequence is memoryless in the multi-session-stitched sense'', which is the appropriate null for the primary $\Cmu$ comparison but conflates within- and inter-session contributions to the rejected null.

To probe within-session structure specifically we generate within-session-shuffle surrogates that permute waiting times independently within each session, preserving the per-session waiting-time multiset (and therefore the per-session marginal distribution) while destroying only within-session temporal ordering. Statistics computed against this null reflect what we would see if the source had no within-session memory but the same activity-rate regimes across sessions. We apply three within-session-shuffle tests at $L = 5$ with 5000 surrogate draws each:

\paragraph{Within-session ACF.} The autocorrelation function of the raw waiting-time sequence (Section~\ref{sec:emachine_results}) restricted to within-session pairs (lags 1--20). For \frba{}, only 1 of 20 lags exceeds the 95th percentile of the surrogate distribution: under a null that preserves session-level marginal differences, the within-session ACF is consistent with no temporal structure. The 19/20-lags-significant result of Section~\ref{sec:emachine_results} was therefore measuring the combined within- and inter-session ACF, with the persistence dominated by inter-session structure. For \frbb{}, 2 of 20 lags exceed the surrogate band, also consistent with no within-session ACF structure (vs.\ 20/20 against the permutation null).

\paragraph{Within-session lag-1 MI.} The lag-1 mutual information of the symbolised sequence (Section~\ref{sec:mi_diagnostic}) computed on within-session symbol pairs only. For \frba{}, only the $k = 5$ result is borderline ($p_\mathrm{raw} = 0.043$, $p_\mathrm{adj} = 0.170$ across the $k$-sweep with $m = 4$); the $k \leq 4$ results are unambiguously null. For \frbb{}, no $k$ value yields $p_\mathrm{raw} < 0.27$.

\paragraph{Boundary-free $\Cmu$.} The boundary-free reconstruction of Section~\ref{sec:boundary_free_method} repeated for each of the 5000 within-session-shuffle surrogates; results in Table~\ref{tab:boundary_free_surr}. For \frba{}, the $k = 5$ boundary-free $\Cmu = 1.070\bits$ exceeds the surrogate distribution at $p_\mathrm{raw} = 0.033$ but does not survive Benjamini--Hochberg correction across the $k$-sweep ($p_\mathrm{adj} = 0.134$). At $k \leq 4$ the boundary-free real $\Cmu = 0$ lies below the surrogate mean ($\sim 0.2\bits$), which reflects the noise floor of CSSR's $\chi^2$ state-splitting test at finite $N$ rather than physical structure in the surrogate; the test for \frba{} at $k \leq 4$ therefore has no power to detect signal weaker than the noise floor. For \frbb{}, the boundary-free $\Cmu = 0.985\bits$ at $k = 4$ does not exceed the surrogate mean ($\Cmu^\mathrm{surr} = 0.45\bits$; $p_\mathrm{raw} = 0.130$), consistent with the source's signal arising entirely from inter-session structure that within-session shuffling preserves.

\begin{table}
    \centering
    \begin{tabular}{@{}llcccc@{}}
        \toprule
        Source & $k$ & $\Cmu^\mathrm{real}$ & $\Cmu^\mathrm{surr}$ & $p_\mathrm{raw}$ & $p_\mathrm{adj}$ \\
        \midrule
        \frba & 2 & 0.000 & $0.23 \pm 0.43$ & 0.262 & 0.349 \\
              & 3 & 0.000 & $0.18 \pm 0.37$ & 0.192 & 0.349 \\
              & 4 & 0.000 & $0.21 \pm 0.38$ & 0.416 & 0.416 \\
              & 5 & 1.070 & $0.25 \pm 0.38$ & \textbf{0.033} & 0.134 \\
        \frbb & 2 & 0.000 & $0.20 \pm 0.39$ & 0.388 & 0.409 \\
              & 3 & 0.000 & $0.28 \pm 0.44$ & 0.409 & 0.409 \\
              & 4 & 0.985 & $0.45 \pm 0.48$ & 0.130 & 0.409 \\
              & 5 & 0.895 & $0.56 \pm 0.49$ & 0.333 & 0.409 \\
        \bottomrule
    \end{tabular}
    \caption{Boundary-free $\Cmu$ (bits) versus within-session-shuffle surrogates (5000 draws per cell) at $L = 5$, $\alpha = 0.001$. $\Cmu^\mathrm{surr}$ entries report mean $\pm$ standard deviation across surrogates. $p_\mathrm{raw}$ is the one-sided upper-tail $p$-value; $p_\mathrm{adj}$ is Benjamini--Hochberg-adjusted across the four $k$ rows per source ($m = 4$). Bold marks $p_\mathrm{raw} < 0.05$; no entry survives BH correction at the 5\% FDR threshold.}
    \label{tab:boundary_free_surr}
\end{table}

\subsection{Interpretation}
\label{sec:boundary_free_interp}

The verdicts of these tests are reported and combined with the session-boundary controls in Section~\ref{sec:one_bit}; here we record only the two points specific to the boundary-free machinery. First, for \frbb{} the within-session-shuffle surrogates of the boundary-free reconstruction themselves yield $\Cmu \sim 0.45\bits$ because they preserve the inter-session marginal contrasts that drive the signal; the surrogate baseline, not just the real value, reflects inter-session structure. Second, whether \frba{}'s borderline $k = 5$ within-session signal ($p_\mathrm{raw} \approx 0.03$--$0.04$ in two independent statistics, neither surviving multiple-comparison correction) is genuinely absent or merely below the detection threshold at $N_\mathrm{wt} \lesssim 121$ per session cannot be determined from current data; resolving the ambiguity requires longer continuous sessions. In both cases the multi-session $\Cmu$ of Section~\ref{sec:emachine_results} remains a valid descriptor of the concatenated sequence as observed; the boundary-free analysis clarifies the timescale on which that memory lives.

\section{CSSR Hyperparameter and Symbolisation-Scheme Sensitivity}
\label{app:hyperparameters}

The $\Cmu$ values reported in the main text rest on several discretionary choices in the analysis pipeline. This appendix quantifies how sensitive the results are to three of them: the multiple-comparison framing applied across the $k$-sweep, the CSSR hyperparameters $L$ and $\alpha$, and the symbolisation scheme that maps waiting times to symbols.

\paragraph{Multiple-comparison framing.} The two FDR framings reported in Table~\ref{tab:surrogates} embody different views of the hypothesis family. The per-source correction ($m = 4$ tests per source) treats the three FRB sources as independent astronomical objects, which is defensible given their disparate locations and host environments. The global correction ($m = 12$) treats all source--$k$ combinations as a single family. Because the $k$-sweep is a robustness check on the same temporal structure rather than 12 independent hypotheses, neither framing is unambiguously correct; the primary claim in the main text rests on the per-source framing, with the global numbers reported as a more conservative sensitivity check.

\paragraph{CSSR hyperparameters.} $\Cmu$ is insensitive to the significance level $\alpha$ of the state-splitting test across the range $\alpha = 0.01$ to $0.0001$, but depends on the maximum history length $L$: \frba{} requires the full $L = 5$ to detect its $\Cmu$, while \frbb{} stabilises from $L = 4$. Because $L = 5$ is the largest history length we sweep, and the CSSR estimate is an $L$-truncated lower bound, behaviour at $L > 5$ is untested; the reported values should be read accordingly.

\paragraph{Symbolisation scheme.} Quantile discretisation (our primary method) yields $\Cmu = 1.116 / 0.986 / 0.000\bits$ for \frba{} / \frbb{} / \frbc{} respectively at $k = 4$. Equal-width binning yields $\Cmu = 0$ for all three sources, which is expected given the extreme bin imbalance produced by equal-width partitioning of heavy-tailed waiting-time distributions (most waiting times fall in a single bin, destroying temporal contrast). Log-spaced binning yields $\Cmu = 1.014 / 0.000 / 0.000\bits$: \frba's signal is partially recovered, while \frbb's vanishes. The log-spaced result for \frbb{} is consistent with the inter-session origin of its memory: log-spacing, like per-session quantile boundaries, alters the mapping from waiting-time values to symbols in a way that can remove the inter-session contrast. The memory detection for \frbb{} is therefore not robust across binning schemes; it requires quantile discretisation, which by construction encodes distributional differences between sessions. The \frba{} detection survives under log-spacing ($\Cmu = 1.014\bits$), reflecting the same multi-session structure that the canonical quantile result captures; the boundary-free analysis of Appendix~\ref{app:boundary_free} shows this multi-session structure is itself predominantly inter-session in origin. The sensitivity to symbolisation scheme indicates that the surrogate tests, which use the same binning as the real data, provide the most reliable inference; the absolute value of $\Cmu$ should be interpreted in the context of the chosen discretisation.

\section{The CSSR Algorithm}
\label{app:cssr}

The CSSR algorithm \citep{Shalizi2004, Shalizi2002} infers the \emachine{} from a single observed symbol sequence. The algorithm proceeds in three stages:

\begin{enumerate}
    \item \textbf{Suffix tree construction.} For each history of length $\ell = 0, 1, \ldots, L$ (where $L$ is a user-specified maximum), estimate the conditional distribution $\hat{P}(X_{t+1} \mid \overleftarrow{x}_\ell)$ over next symbols given the observed $\ell$-history.
    \item \textbf{State splitting.} Starting from a single state containing all histories of a given length, iteratively test whether two histories should be split into different causal states. The splitting criterion is a $\chi^2$ homogeneity test at significance level $\alpha$: if the conditional distributions of two histories differ at significance $\alpha$, they are assigned to different states.
    \item \textbf{Transition matrix construction.} Once the causal states have been identified, estimate the transition probabilities $P(x \mid s)$ and the deterministic transition function $\delta(s, x)$ from the data.
\end{enumerate}

The two hyperparameters are the maximum history length $L$ and the significance level $\alpha$. Larger $L$ allows detection of longer-range dependencies but requires more data; smaller $\alpha$ is more conservative (merges more states) and guards against over-splitting in small samples; their sensitivity is examined in Appendix~\ref{app:hyperparameters}.

The following pseudocode summarises the operational procedure used in this work. The algorithm takes as input a symbol sequence $\bm{x} = x_1 x_2 \ldots x_N$, a maximum history length $L$, and a significance level $\alpha$.

\begin{enumerate}
    \item \textbf{Initialise.} For each history $w$ of length $\ell \leq 1$ (the empty history and each single symbol), estimate the conditional distribution $\hat{P}(X_{t+1} \mid x_{t-\ell+1} \ldots x_t = w)$ by counting.
    \item \textbf{For $\ell = 1$ to $L$:}
    \begin{enumerate}
        \item For each pair of histories $(w, w')$ of length $\ell$ currently in the same state, perform a $\chi^2$ test of homogeneity on their conditional distributions.
        \item If the test rejects at level $\alpha$, split $w$ and $w'$ into separate states.
        \item Extend all histories by one symbol (append each $a \in \mathcal{A}$) and repeat.
    \end{enumerate}
    \item \textbf{Determinise.} After reaching $\ell = L$, split any state whose histories do not all transition to a single successor state under a given next symbol, so that the resulting transition structure is deterministic (each state and symbol yield a unique successor).
    \item \textbf{Construct.} Build the transition matrices $T^{(a)}_{ij} = P(S_{t+1} = j, X_{t+1} = a \mid S_t = i)$ from the assigned state labels. (This emit-on-transition bookkeeping is equivalent, for a unifilar machine, to the emit-on-state convention $P(X_t \mid S_t)$ of Section~\ref{sec:emachines}: the emitted symbol and current state jointly fix the successor.)
    \item \textbf{Output.} Return the \emachine: states, transitions, stationary distribution, $\Cmu$, $\hmu$.
\end{enumerate}

\section{IAAFT Surrogate Decomposition of Linear and Nonlinear Memory}
\label{app:iaaft}

The permutation surrogates used in Section~\ref{sec:emachine_results} destroy all temporal structure, including linear autocorrelation. Since both FAST sources show significant autocorrelation (Section~\ref{sec:emachine_results}), the elevated $\Cmu$ could merely reflect the known ACF structure rather than indicating higher-order (nonlinear) temporal dependence. To test this, we generated 1000 IAAFT surrogates \citep{SchreiberSchmitz1996} for each source and $k$. IAAFT surrogates preserve both the marginal amplitude distribution and the power spectrum (and hence the linear autocorrelation) of the original series while destroying higher-order temporal structure. Because each IAAFT iteration rescales the surrogate to the rank-ordered amplitudes of the observed series, every surrogate is a reordering of the actual waiting times, so positivity and the heavy-tailed marginal are preserved exactly and no parametric assumption about the waiting-time distribution enters; only the temporal arrangement compatible with the matched power spectrum is randomised. If $\Cmu$ is significant against IAAFT surrogates, the temporal memory extends beyond what linear correlations capture.

Table~\ref{tab:iaaft} shows the results. The two FAST sources show qualitatively different behaviour. \frba{} yields $\Cmu$ marginally above the IAAFT surrogate distribution at $k = 4$ ($p = 0.032$, $z = 2.3$), with a borderline result at $k = 5$ ($p = 0.076$). This suggests that \frba{} possesses temporal structure beyond its (weak) linear autocorrelation (this $p$-value is not corrected for multiple testing across $k$ and should be regarded as suggestive). In contrast, \frbb{} does not significantly exceed the IAAFT surrogates at any $k$ (lowest $p = 0.098$ at $k = 4$), indicating that its strong linear autocorrelation (ACF(1) = 0.247; Section~\ref{sec:emachine_results}) accounts for most of the detected $\Cmu$. \frbc{} remains consistent with the IAAFT null at all $k$, as expected.

The IAAFT analysis does not alter the measured $\Cmu$ values; both FAST sources still carry $\Cmu$ of order one bit. The IAAFT test decomposes the origin of that memory: the temporal memory detected in \frba{} has a suggestive nonlinear component, while the memory in \frbb{} is consistent with being primarily linear. In both cases, the waiting-time process requires $\sim$1\,bit of predictive information, but the nature of the temporal dependence appears to differ between sources.

\begin{table*}
    \centering
    \begin{tabular}{@{}lcccc@{}}
        \toprule
        Source & $k=2$ & $k=3$ & $k=4$ & $k=5$ \\
        \midrule
        \frba & 0.918 (0.195) & 0.000 (1.000) & 1.116 (\textbf{0.032}) & 0.831 (0.076) \\
        \frbb & 0.000 (0.597) & 0.000 (0.473) & 0.986 (0.098) & 0.900 (0.201) \\
        \frbc & 0.000 (1.000) & 0.000 (1.000) & 0.000 (0.391) & 0.000 (1.000) \\
        \bottomrule
    \end{tabular}
    \caption{IAAFT surrogate testing results. Each entry shows $\Cmu^\mathrm{real}$ (bits) with the IAAFT surrogate $p$-value in parentheses; bold entries indicate $p < 0.05$. IAAFT surrogates preserve the marginal distribution and power spectrum (and hence the linear autocorrelation), destroying only higher-order temporal structure, so significance here indicates temporal memory beyond linear autocorrelation.}
    \label{tab:iaaft}
\end{table*}

\section{Robustness Diagnostics}
\label{app:robustness}

This appendix collects the quantitative detail behind the robustness checks summarised in Section~\ref{sec:robustness}: block-bootstrap confidence intervals, the behaviour of $\Cmu$ across alphabet size, and the mutual-information diagnostic.

\paragraph{Bootstrap confidence intervals.} Table~\ref{tab:bootstrap} reports the 95\% block-bootstrap confidence intervals (CIs) for $\Cmu$ (Section~\ref{sec:bootstrap}). The intervals are wide because $\Cmu$ is bimodal across bootstrap replicates: individual replicates yield either $\Cmu \approx 0$ or $\Cmu \approx 1$, with little intermediate density. This bimodality is consistent with CSSR's $\chi^2$ state-splitting at finite $N$, where either the test fires (multiple causal states, $\Cmu \approx 1\bits$) or it does not (single state, $\Cmu = 0$), rather than with genuine bimodality of the underlying parameter. The synthetic minimum-$N$ analysis of Appendix~\ref{app:validation} confirms this directly: a fraction of \emph{known-structured} HMM realisations collapse to a single-state $\Cmu = 0$ at finite $N$ (falling from ${\sim}0.5$--$0.75$ at $k = 2$ to ${\sim}0.1$ at $k = 4$ and essentially independent of $N$), so an individual resample returning $\Cmu = 0$ is an expected algorithmic outcome on data that genuinely carry structure rather than evidence that the signal is absent. The bootstrap and surrogate tests answer different questions. Bootstrap CIs quantify estimation uncertainty: ``how precisely is $\Cmu$ measured?'' The surrogate test is a hypothesis test: ``is the observed $\Cmu$ consistent with a memoryless process?'' The hypothesis test rejects memorylessness unambiguously ($p \leq 0.01$), even though the bootstrap intervals are wide; the wide CIs do not independently corroborate the detection but quantify how imprecisely the absolute value of $\Cmu$ is determined. A further design caveat applies: the selected block length of 6--12 waiting times (Section~\ref{sec:bootstrap}) is far shorter than a typical session, so resampling scrambles precisely the session-level structure that Section~\ref{sec:one_bit} identifies as the carrier of the signal. The wide intervals and zero medians at some $k$ therefore partly reflect the resampling scale itself rather than only CSSR's finite-$N$ bimodality, and a session-level block scheme (resampling whole sessions) would be the natural refinement. To check sensitivity to the block-length choice directly, we re-ran the full bootstrap at half and double the selected block length for every source and $k$ (block lengths 3--24). The intervals respond as expected for dependent data: upper limits shrink by up to ${\sim}30$ per cent at half the block length and widen by up to ${\sim}35$ per cent at double. The lower limit is zero in every configuration, so the interpretation above holds at all tested block lengths: the intervals are wide, include zero, and quantify estimation uncertainty without independently corroborating the detection.

\begin{table*}
    \centering
    \begin{tabular}{@{}lcccc@{}}
        \toprule
        Source & $k=2$ & $k=3$ & $k=4$ & $k=5$ \\
        \midrule
        \frba & 0.918 (0.000) [0.000, 0.993] & 0.000 (0.000) [0.000, 1.377] & 1.116 (0.679) [0.000, 1.623] & 0.831 (0.815) [0.000, 1.846] \\
        \frbb & 0.000 (0.000) [0.000, 1.442] & 0.000 (0.000) [0.000, 1.373] & 0.986 (0.773) [0.000, 1.658] & 0.900 (0.850) [0.000, 1.832] \\
        \frbc & 0.000 (0.000) [0.000, 0.920] & 0.000 (0.000) [0.000, 0.998] & 0.000 (0.000) [0.000, 1.417] & 0.000 (0.000) [0.000, 1.341] \\
        \bottomrule
    \end{tabular}
    \caption{Block-bootstrap 95\% confidence intervals for $\Cmu$ (bits) across sources and $k$. Each cell shows $\Cmu^\mathrm{real}$ (bootstrap median) [2.5th, 97.5th percentile] over 200 replicates. The bootstrap distribution is bimodal: individual replicates yield either $\Cmu \approx 0$ or $\Cmu \approx 1$, reflecting whether the resample preserves sufficient temporal structure for CSSR to detect. At $k = 4$, the median is non-zero for both FAST sources (0.679 and 0.773\,bits) but zero for \frbc.}
    \label{tab:bootstrap}
\end{table*}

\begin{figure}
    \centering
    \includegraphics[width=\columnwidth]{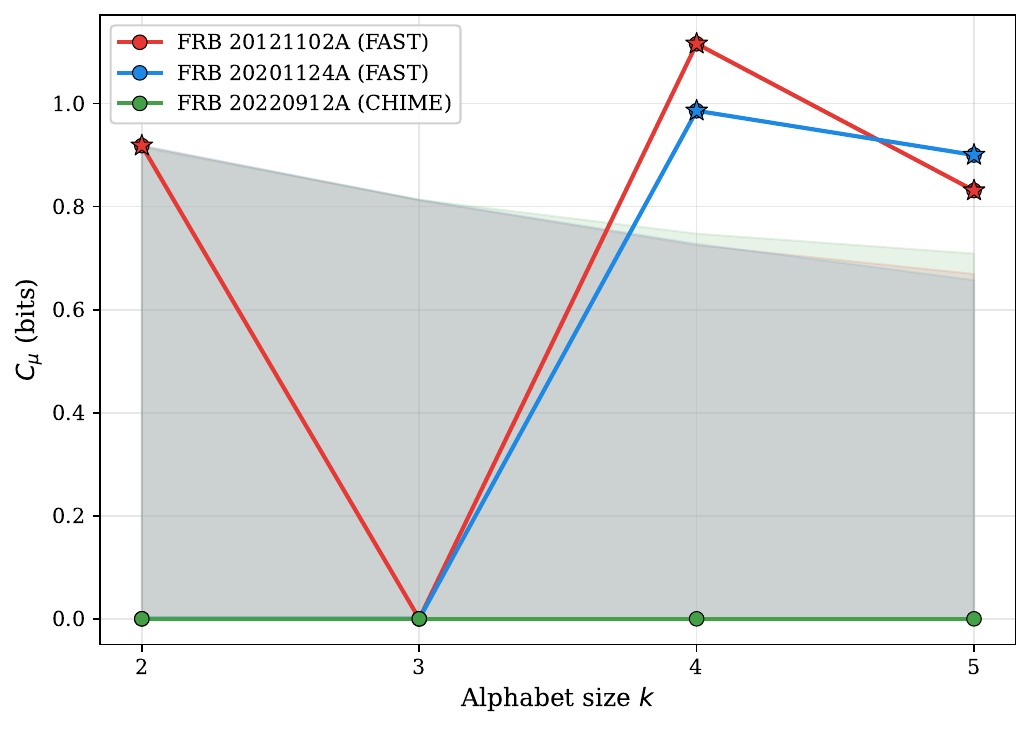}
    \caption{Statistical complexity $\Cmu$ as a function of alphabet size $k$. Solid lines show the real $\Cmu$ for each source; shaded bands show the per-source 5th--95th percentile of the corresponding permutation surrogate distribution (the three bands overlap closely because surrogate $\Cmu$ distributions are similar across sources). Stars mark points where $p < 0.05$; uncertainties on the real-data points are quantified by the block-bootstrap intervals of Table~\ref{tab:bootstrap}. For sources with bimodal surrogate $\Cmu$ distributions (a pile-up at zero with a secondary cluster near 0.7--0.8\,bits, as in Figure~\ref{fig:surrogates}), the upper edge of the band can sit close to the real $\Cmu$ value while the tabulated $p$-value remains $< 0.05$. The drop to $\Cmu = 0$ between $k = 2$ and $k = 3$ for \frba{} is a CSSR coarse-resolution failure, not a genuine non-monotonic feature: at $k = 2$ the contingency-table $\chi^2$ test fires reliably; at $k = 3$ expected cell counts fall below the threshold and the algorithm cannot reject the single-state null. The MI diagnostic of Figure~\ref{fig:mi} confirms that genuine temporal structure is present at $k = 3$.}
    \label{fig:cmu_vs_k}
\end{figure}

\paragraph{Alphabet-size sweep.} Figure~\ref{fig:cmu_vs_k} shows $\Cmu$ as a function of $k$ for all three sources, with shaded bands indicating the 5th--95th percentile range of the surrogate distribution. The comparative ranking is robust across all $k$: \frbc{} yields $\Cmu = 0$ at every resolution, while the FAST sources exceed it wherever CSSR detects structure. At $k = 2$ and $k = 3$, the FAST sources show $\Cmu = 0$ for some configurations; this is a CSSR reconstruction artefact at coarse resolution, not absence of temporal structure, as the mutual-information diagnostic below establishes.

\begin{figure*}
    \centering
    \includegraphics[width=\textwidth]{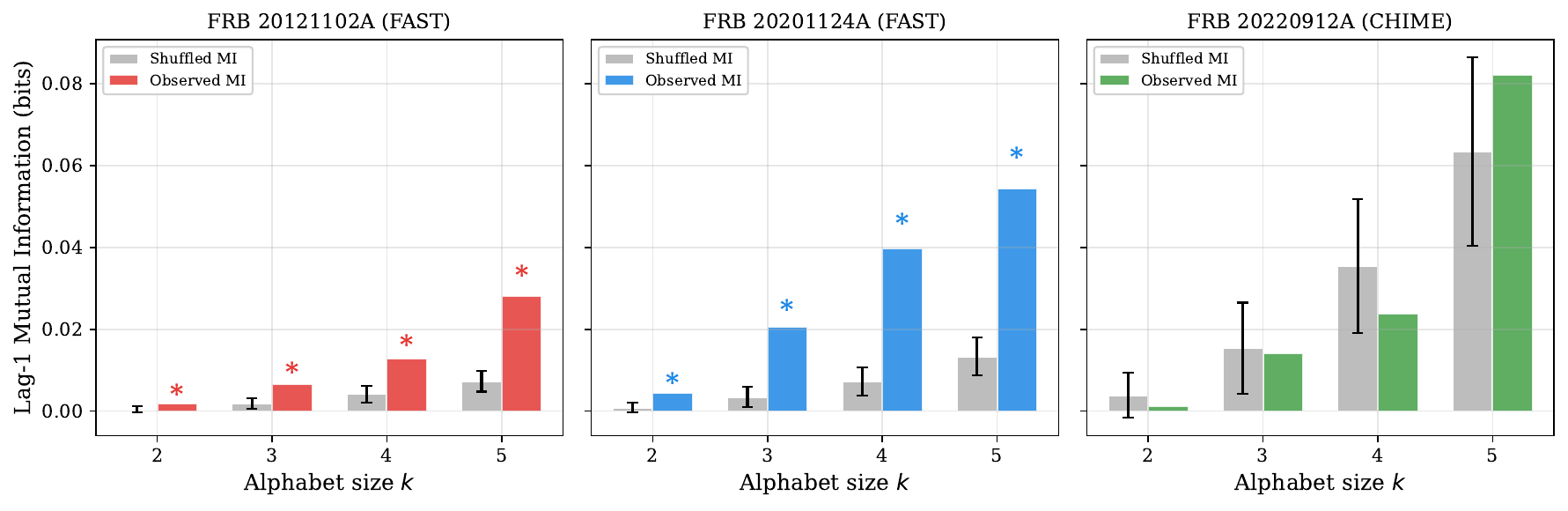}
    \caption{Lag-1 mutual information of the symbolised sequence (coloured bars) compared to 1000 shuffled surrogates (grey bars, with error bars showing $\pm 1\sigma$ of the shuffled distribution). Stars indicate $p < 0.05$. Temporal structure is preserved in the symbols at every tested $k$ for both FAST sources; \frbc{} is consistent with IID at all $k$. For \frbc{} at $k = 5$ the observed MI ($0.082$\,bits) is the highest of any $k$ for this source but remains within the shuffled distribution ($p = 0.190$); the bar appears comparable to the shuffled $\pm 1\sigma$ envelope by construction, since $\sigma$ corresponds to the $\sim$84th percentile of an approximately Gaussian shuffled distribution.}
    \label{fig:mi}
\end{figure*}

\paragraph{Mutual-information diagnostic.} Figure~\ref{fig:mi} shows the lag-1 MI of the symbolised sequence (Section~\ref{sec:mi_diagnostic}) compared against shuffled surrogates. For both FAST sources, the lag-1 MI is significantly above the shuffled baseline at every tested resolution, confirming that the symbolisation preserves temporal structure even at the coarsest binning (at $k = 2$, \frba{} $p = 0.032$ and \frbb{} $p = 0.026$). The fact that $\Cmu = 0$ at $k = 2, 3$ for some FAST configurations, despite significant MI, indicates that CSSR fails to detect the structure at coarse resolutions where conditional transition probabilities become nearly uniform: a reconstruction failure, not an absence of signal. Appendix~\ref{app:validation} shows the same collapse on synthetic processes of known structure, so this coarse-resolution null is an expected property of CSSR rather than a source-specific result. For \frbc, the MI is not significant at any $k$ ($p > 0.1$ in all cases); the symbol sequence is indistinguishable from an IID process at every resolution, consistent with the $\Cmu = 0$ result.

\section{Robustness to Completeness Cuts}
\label{app:fluence_control}

Because the detected memory is entirely inter-session (Section~\ref{sec:one_bit}), any instrumental property that varies from session to session is in principle degenerate with intrinsic regime switching (Section~\ref{sec:limitations}). This appendix tests the most accessible form of that degeneracy: structure carried by bursts near the detection limit, whose recovery is most sensitive to session-to-session threshold variation. We repeat the detection analysis on completeness-cut subsamples, removing all bursts with fluence below a threshold $F_{\min}$ so that every retained burst is bright enough to be detected irrespective of the instrument's session-to-session sensitivity. The thresholds are anchored at each source's published 90 per cent completeness value, $F_{90} = 0.02$\,Jy\,ms for \frba{} \citep{Li2021} and $E_\mathrm{th} = 3.6\times10^{36}$\,erg, equivalent to ${\approx}0.032$\,Jy\,ms, for \frbb{} \citep{Zhang2022}, supplemented by multiples of the anchor and by the 10th, 25th and 50th percentiles of each catalogue's fluence distribution. Observing sessions are kept fixed at those of the full catalogue, since they reflect the telescope schedule rather than the burst sample; the intra-session waiting times of the retained bursts are recomputed, symbolised with their own quantile boundaries, and passed through the same reconstruction and 1000-draw permutation-surrogate test as the real data (Section~\ref{sec:surrogates}), together with the session-shuffle test at $k = 4$. A no-cut rung reproduces the canonical values of Table~\ref{tab:surrogates}.

A null result from a fluence cut is interpretable only against a power reference: removing a burst merges its two adjacent waiting times into one, so any depletion, selective or not, perturbs the symbol sequence directly. For each rung we therefore draw 200 uniform random thinnings of the full burst list to the same number of bursts as the cut and pass each through the identical pipeline at $k = 4$. The fraction of thinned realisations retaining $\Cmu \geq 0.5\bits$, reported as $f_{0.5}$ in Table~\ref{tab:fluence_control}, estimates the probability that a detection-scale $\Cmu$ survives the removal of the same number of bursts chosen at random. The control is asymmetric by construction: a detection that survives the cut is strong evidence against an instrumental origin, whereas a joint collapse of cut and baseline is uninformative. A well-powered collapse under the cut alone would itself admit two readings, instrumental contamination or genuine memory carried disproportionately by faint bursts, so even the negative branch would not uniquely indicate systematics.

\begin{table*}
    \centering
    \begin{tabular}{@{}llrrcccc@{}}
        \toprule
        Source & Cut & $F_{\min}$ (Jy ms) & $N_\mathrm{burst}$ & $\Cmu^{k=4}$ ($p$) & $\Cmu^{k=5}$ ($p$) & $p_\mathrm{shuffle}$ & $f_{0.5}$ \\
        \midrule
        \frba & $F_{90}$      & 0.020 & 1580 & 0.000 (0.245) & 0.000 (0.239) & 0.564 & 0.20 \\
              & $P_{10}$      & 0.027 & 1486 & 0.000 (1.000) & 0.000 (0.174) & 1.000 & 0.21 \\
              & $1.5\,F_{90}$ & 0.030 & 1441 & 0.767 (\textbf{0.028}) & 0.000 (1.000) & 0.129 & 0.27 \\
              & $2\,F_{90}$   & 0.040 & 1276 & 0.000 (1.000) & 0.000 (0.250) & 1.000 & 0.30 \\
              & $P_{25}$      & 0.042 & 1241 & 0.000 (0.220) & 0.767 (\textbf{0.028}) & 0.228 & 0.34 \\
              & $P_{50}$      & 0.065 &  826 & 0.000 (1.000) & 0.000 (1.000) & 1.000 & 0.29 \\
        \addlinespace
        \frbb & $F_{90}$      & 0.032 &  823 & 0.679 (0.120) & 1.000 (\textbf{0.008}) & 0.307 & 0.78 \\
              & $P_{10}$      & 0.040 &  794 & 0.965 (\textbf{0.012}) & 0.581 (0.177) & 0.653 & 0.71 \\
              & $1.5\,F_{90}$ & 0.048 &  757 & 0.816 (\textbf{0.017}) & 0.591 (0.170) & 1.000 & 0.76 \\
              & $2\,F_{90}$   & 0.064 &  697 & 1.376 (\textbf{0.001}) & 1.000 (\textbf{0.011}) & 0.545 & 0.65 \\
              & $P_{25}$      & 0.073 &  663 & 1.000 (\textbf{0.004}) & 1.000 (\textbf{0.007}) & 0.505 & 0.66 \\
              & $P_{50}$      & 0.221 &  441 & 0.000 (1.000) & 0.000 (1.000) & 1.000 & 0.34 \\
        \bottomrule
    \end{tabular}
    \caption{Completeness-cut control. For each rung, all bursts with fluence below $F_{\min}$ are removed, sessions are kept fixed at those of the full catalogue, and the detection analysis is repeated: $\Cmu$ with its 1000-draw permutation-surrogate $p$-value at $k = 4$ and $k = 5$, and the session-shuffle $p$-value at $k = 4$ (100 permutations). All $p$-values use the $(r+1)/(n+1)$ Monte-Carlo estimator \citep{PhipsonSmyth2010}. $f_{0.5}$ is the power reference: the fraction of 200 random thinnings of the full burst list to the same $N_\mathrm{burst}$ whose $k = 4$ reconstruction retains $\Cmu \geq 0.5\bits$. Anchors $F_{90}$ are the published 90 per cent completeness thresholds (\frba{}: 0.02\,Jy\,ms, \citealt{Li2021}; \frbb{}: $E_\mathrm{th} = 3.6\times10^{36}$\,erg ${\approx}\,0.032$\,Jy\,ms, \citealt{Zhang2022}); $P_{10}$, $P_{25}$, $P_{50}$ are percentiles of each catalogue's fluence distribution. Bold marks $p < 0.05$. Entries displayed as $0.000$ can be non-zero at machine precision (cf.\ Table~\ref{tab:surrogates}). \frbc{} remains null at every rung and is omitted.}
    \label{tab:fluence_control}
\end{table*}

For \frbb{} the detection survives the completeness cut and well beyond it. At $F_{90}$ the $k = 5$ detection holds ($\Cmu = 1.000\bits$, $p = 0.008$); at every rung up to $2\,F_{90}$ and $P_{25}$ a detection persists at one or both of the anchor resolutions, and at $2\,F_{90}$, with 21 per cent of bursts removed, the $k = 4$ result is the strongest in the table ($p = 0.001$). The detection fails only at the 50th-percentile cut, where the thinning baseline collapses comparably ($f_{0.5} = 0.34$): a loss of power, not of signal. \frbb{}'s inter-session signal is therefore not carried by bursts near the detection limit.

For \frba{} the cut at $F_{90}$ removes only 72 of 1652 bursts (4.4 per cent) yet collapses the $k = 4$ detection ($\Cmu = 0$, $p = 0.245$) and the session-shuffle signal ($p = 0.564$). The thinning baseline shows, however, that this is what almost any removal of 72 bursts does: only 20 per cent of matched random thinnings retain a detection-scale $\Cmu$, and across all rungs the fluence-cut outcome is statistically typical of the thinning distribution (the fraction of thinnings doing as badly as or worse than the cut at $k = 4$ ranges over 0.66--0.83). The isolated revivals at single resolutions ($\Cmu = 0.767$ at $k = 4$ for $1.5\,F_{90}$ and at $k = 5$ for $P_{25}$) mirror the scattered non-zero entries of the windowing test (Section~\ref{sec:bias_results}) and are consistent with the binary firing behaviour of CSSR's state-splitting test near threshold (Appendix~\ref{app:robustness}). The control is therefore power-limited for this source: it provides no evidence for a sensitivity-correlated component, but cannot exclude one either.

A complementary covariate check asks whether session-to-session sensitivity variation correlates with the timing structure at all. For each session we take the minimum detected fluence as a proxy for that session's effective detection threshold and correlate it with the session's median waiting time and mean symbol. The raw correlations for \frba{} are substantial (Spearman $\rho = 0.66$ and $0.72$ over the 38 of its 39 sessions that contain at least two bursts), but the minimum is an order statistic: sessions with more bursts sample deeper into the fluence distribution, which induces exactly this correlation under perfectly constant sensitivity. Calibrated against a permutation null that shuffles fluences across bursts while keeping arrival times and session membership fixed (null $\rho = 0.54 \pm 0.09$), the observed values are not significant ($p_\mathrm{perm} = 0.18$ and $0.056$). For \frbc{} the same check is null ($\rho \approx 0.03$, $p_\mathrm{perm} \approx 0.4$); \frbb{}, with four sessions, has too few for the correlation to be meaningful.

\section{Reconciliation with Prior FRB Temporal-Analysis Literature}
\label{app:literature}

\paragraph{Scalar complexity measures.} \citet{Zhang2024timeenergy} applied the Pincus index (approximate entropy) and maximum Lyapunov exponent to FRB burst arrival times and energies, finding that repeating FRBs behave more randomly and less chaotically than pulsars, earthquakes, or solar flares. \citet{SangLin2024} subsequently applied the Pincus index, the Hurst exponent, and non-Gaussian fluctuation statistics to FAST repeater data, reaching the related conclusion that the waiting-time sequences exhibit long-term memory and scale-invariant behaviour consistent with self-organised criticality. Our analysis measures a different quantity: $\Cmu$ quantifies predictive memory, while approximate entropy measures regularity and the Lyapunov exponent measures dynamical instability. These are complementary perspectives on the same underlying process. Our measurement of $\Cmu \approx 1\bits$ for the FAST sources indicates that the process, while appearing random by scalar complexity measures, possesses genuine predictive structure that these measures do not capture. The numerical scalar measures we computed for our three sources are tabulated in Appendix~\ref{app:scalar}.

\paragraph{Long-range correlation and ``memory'' analyses.} Several studies report ``memory'' or scale-invariant structure in repeater waiting times: \citet{Du2024scaling} find universal scaling reminiscent of solar flares, and \citet{Wang2024memory} report conditional dependence of waiting times on the preceding interval, both consistent with the self-organised-criticality framing of \citet{SangLin2024}. \citet{TotaniTsuzuki2023} provide the most direct evidence for short-timescale dependence: a correlation-function analysis in the two-dimensional time--energy space of nearly 7000 bursts from the same three sources analysed here finds that each burst has a 10--60 per cent chance of triggering an aftershock, at a rate decaying with separation as the Omori--Utsu power law of earthquakes, a signal concentrated at the shortest separations and stable across changes in source activity. Because our pipeline symbolises waiting times into quantile bins that are logarithmically broad for these heavy-tailed distributions, an enhancement confined to separations within the lowest bin's dynamic range only weakly alters which symbol follows which; this is the likely reason aftershock-like clustering does not register in the within-session controls of Appendix~\ref{app:boundary_free}, and a quantitative reconciliation under a common burst definition remains open (Section~\ref{sec:interpretation}). \citet{WangWuDai2023} more specifically report memory in \frba{} and \frbb{} on minute-to-hour (i.e.\ within-session) timescales, detected via the Hurst exponent and growth in burst-rate structure. This is a long-range-correlation or persistence statistic, a different object from the predictive complexity $\Cmu$: a sequence can exhibit Hurst-detectable persistence without a finite-state predictive model surviving at our symbolisation, and their signal lies precisely on the sub-hour scales where our within-session predictive-memory test is null or power-limited (Section~\ref{sec:one_bit}). Whether such within-session persistence reflects predictive structure that survives our boundary-free and within-session-shuffle controls is left to future work.

\paragraph{Parametric waiting-time fits.} A series of studies have fitted parametric waiting-time distributions to FRB~20121102A bursts: \citet{Oppermann2018} adopted a Weibull form, \citet{Gourdji2019} and \citet{Aggarwal2021} both reported log-normal distributions (centred at ${\sim}200$\,s and ${\sim}75$\,s respectively, the latter from a larger Arecibo sample), \citet{Cruces2021} reported departures from exponentiality consistent with non-Poisson emission, and \citet{Jahns2023} described the November 2018 burst storm as a Poisson process with time-varying rate (between $0$ and $218$ bursts per hour). At first glance, these results, particularly the finding that waiting times within active windows are consistent with Poisson statistics, may appear to contradict our detection of temporal memory, but the two are compatible. A process that alternates between two Poisson rates produces individual waiting times that are each exponentially distributed (conditioned on the current state), yet the \emph{sequence} of waiting times carries predictive information: observing several short waiting times in succession increases the posterior probability of being in the fast-rate state, and hence of the next waiting time also being short. Such a process has $\Cmu > 0$ (encoding which rate is currently active) despite having Poisson marginal statistics within each state. Our $\Cmu$ measurement is sensitive to this kind of sequential dependence, which marginal distribution fitting cannot detect. That $\Cmu > 0$ after symbolisation shows that the temporal \emph{ordering} of waiting times, not just their distribution, carries information, as expected if a hidden state modulates the burst rate. The rate-switching comparison in Section~\ref{sec:literature} shows further that the observed $\Cmu$ exceeds what a fitted two-state Markov-modulated Poisson process alone would produce.

\paragraph{Periodicity searches.} The periodicity analyses of \citet{CHIME2020periodic} and \citet{Rajwade2020} identify periodic activity windows on timescales of days to weeks. These are distinct from the temporal structure we detect in the concatenated waiting-time sequence, which reflects non-periodic session-to-session variability in the source's activity rate. \emachines{} detect structure that need not be periodic; the hidden state transitions may be stochastic, irregular, and non-repeating.

\section{Comparison with Scalar Complexity Measures}
\label{app:scalar}

To contextualise the \emachine{} results, we compute three scalar complexity measures on the raw waiting-time sequences: approximate entropy \citep{Pincus1991}, as applied to FRB timing by \citet{SangLin2024} and \citet{Zhang2024timeenergy}; the maximum Lyapunov exponent, which quantifies dynamical instability; and the multifractal detrended fluctuation analysis \citep[MFDFA;][]{Kantelhardt2002} spectrum width $\Delta\alpha$, which characterises multi-scale heterogeneity. Table~\ref{tab:comparison} summarises the results.

\begin{table}
    \centering
    \begin{tabular}{@{}lccc@{}}
        \toprule
        Source & ApEn ($m=2$) & $\lambda_{\max}$ & $\Delta\alpha$ \\
        \midrule
        \frba & 1.086 & 1.909 & 2.084 \\
        \frbb & 0.891 & 1.887 & 1.029 \\
        \frbc & 1.044 & 1.072 & 1.194 \\
        \bottomrule
    \end{tabular}
    \caption{Scalar complexity measures for the three FRB sources, computed on the raw (unsymbolised) waiting-time sequences. Approximate entropy (ApEn, embedding dimension $m = 2$; natural-log base) and the MFDFA spectrum width $\Delta\alpha$ are dimensionless; the maximum Lyapunov exponent $\lambda_{\max}$ is reported in units of inverse waiting-time step (natural-log base). All values are point estimates without uncertainty quantification and should be interpreted as indicative rather than statistically significant; bootstrap intervals on these scalar measures are not reported here because the $\Cmu$ surrogate framework is our primary inferential tool (Section~\ref{sec:surrogates}).}
    \label{tab:comparison}
\end{table}

All three sources yield ApEn values between 0.89 and 1.09, indicating consistently high irregularity, consistent with the finding of \citet{Zhang2024timeenergy} that FRBs are ``more random'' than pulsars by this measure. These scalar measures provide a different perspective from $\Cmu$: ApEn measures distributional regularity, the Lyapunov exponent measures dynamical instability, and MFDFA characterises multi-scale heterogeneity, whereas $\Cmu$ measures \emph{predictive memory}, that is, the minimum information about the past required for optimal prediction. By approximate entropy, \frbc{} (1.044) sits between the two FAST sources (0.891 and 1.086) and so gives no hint of the categorical split that $\Cmu$ draws between them ($\Cmu \approx 1$ vs.\ $\Cmu = 0$). The Lyapunov exponent does separate \frbc{} (1.072, well below the ${\sim}1.9$ of both FAST sources), but as a measure of dynamical instability rather than predictive memory. The widest multifractal spectrum ($\Delta\alpha = 2.08$) belongs to \frba, suggesting the most complex multi-scale structure, while \frbb{} has the lowest ApEn (0.891), consistent with its strongest persistent autocorrelation. The \emachine{} framework contributes a minimal generative model and a precise memory quantification, complementing rather than superseding these scalar characterisations.

\section{Model Selection for the Rate-Switching Null}
\label{app:mmpp_aic}

The rate-switching comparison of Section~\ref{sec:literature} asks whether the observed $\Cmu$ exceeds that generated by a fitted two-state MMPP. A complementary question is whether the MMPP's transition structure is warranted in the first place, or whether a model with the same marginal waiting-time distribution but no temporal dependence fits the data equally well. We address this with an Akaike Information Criterion (AIC) comparison of the fitted MMPP against the IID two-component exponential mixture of Section~\ref{sec:literature} (same marginal distribution, each interval drawn independently), defining $\Delta\mathrm{AIC} = \mathrm{AIC}_\mathrm{MMPP} - \mathrm{AIC}_\mathrm{mix}$ so that negative values favour the MMPP. The $\Delta\mathrm{AIC}$ values are annotated on Figure~\ref{fig:mmpp}.

The two FAST sources differ. For \frba{} the Markov transitions are strongly warranted ($\Delta\mathrm{AIC} = -24.3$): the data support a genuine rate-switching process, so the $\Cmu$ excess reported in Section~\ref{sec:literature} is a step beyond a model that does capture rate-switching. For \frbb{} the transitions are not favoured ($\Delta\mathrm{AIC} = +1.0$): the mixture fits as well as the MMPP, consistent with \frbb{}'s signal being session-level rate heterogeneity rather than an ordered regime sequence (Section~\ref{sec:interpretation}). In both sources the $\Cmu$ excess over the fitted-MMPP baseline holds regardless, so it does not depend on the MMPP being the preferred parametric description of the waiting times.

\label{lastpage}
\end{document}